\documentstyle[amstex,preprint,aps,epsf,floats]{revtex}

\tighten

\begin{document}

\def\Journal#1#2#3#4{{#1} {\bf #2}, #3 (#4)}

\def\NCA{\em Nuovo Cimento}
\def\NIM{\em Nucl. Instrum. Methods}
\def\NIMA{{\em Nucl. Instrum. Methods} A}
\def\NPB{{\em Nucl. Phys.} B}
\def\NPA{{\em Nucl. Phys.} A}
\def\NP{{\em Nucl. Phys.} }
\def\PLB{{\em Phys. Lett.} B}
\def\PRL{\em Phys. Rev. Lett.}
\def\PRD{{\em Phys. Rev.} D}
\def\PRC{{\em Phys. Rev.} C}
\def\PRA{{\em Phys. Rev.} A}
\def\PR{{\em Phys. Rev.} }
\def\PRep{{\em Phys. Rep.}}
\def\ZPC{{\em Z. Phys.} C}
\def\SJP{{\em Sov. Phys. JETP}}
\def\SJNP{{\em Sov. Phys. Nucl. Phys.}}

\def\FBS{{\em Few Body Systems Suppl.}}
\def\IJMP{{\em Int. J. Mod. Phys.} A}
\def\UJP{{\em Ukr. J. of Phys.}}
\def\CJP{{\em Can. J. Phys.}}
\def\SCI{{\em Science} }

\preprint{\vbox{
\hbox{NT@@UW-99-18} }}
\bigskip
\bigskip

\title{Elastic and Inelastic Neutrino-Deuteron Scattering in Effective
Field Theory}   
\author{Malcolm Butler}
\address{Department of Astronomy and Physics, Saint Mary's University\\
Halifax, NS B3H 3C3 Canada\\
{\tt mbutler@@ap.stmarys.ca}}
\author{Jiunn-Wei Chen}
\address{Department of Physics, University of Washington, \\
Seattle, WA 98195-1560 \\
{\tt jwchen@@phys.washington.edu}}
\maketitle

\begin{abstract}
The differential cross-sections for elastic and inelastic neutrino-deuteron
scattering are calculated analytically using
nucleon-nucleon effective field theory. For elastic scattering, the deuteron
axial form factor and the deuteron strange magnetic moment form factor are
computed to next-to-leading order, including two-body currents. For
inelastic scattering, two neutral current processes $\nu d\rightarrow \nu np$%
, $\overline{\nu }d\rightarrow \overline{\nu }np$ and one charged current
process $\overline{\nu }d\rightarrow e^{+}nn$ are computed to
next-to-leading order. These depend on an isovector axial two-body matrix
element whose value is yet to be fixed by experiment. Potential model
calculations by Kubodera {\it et al}.\ and Ying {\it et al}.\ are reproduced
for different values of the two-body matrix element. This implies that
the differences between the two potential model calculations lie in their
treatment of short distance physics.
 The charged current to neutral current $\overline{\nu }d$
cross-section ratio is confirmed to be insensitive to short distance physics,
and the same ratio is obtained by potential models and the calculation presented
here, within 5\%, for incident incident neutrino energies up to 20 MeV. The
two-body matrix element could be fixed using the parity violating
process $\overrightarrow{e}d\rightarrow e np$.
\end{abstract}

\bigskip \vskip3.0cm \leftline{May, 1999\hfil}

\vfill\eject

\section{Introduction}

The Sudbury Neutrino Observatory (SNO) employs inelastic neutrino-deuteron
scattering to study the solar neutrino flux, and seeks the solution to the
solar neutrino problem. The cross-sections for these processes have never
been measured, leading to a reliance on theoretical calculations in the
analysis of their data. To date, fairly sophisticated potential model
calculations\cite{YHH,KN}, employing different approaches to the inclusion
of meson exchange currents have produced results which agree to within
5-10\% of one another up to neutrino energies of 160~MeV. Still, how well
constrained are these calculations? Given the importance of SNO to the
understanding of neutrino physics, these cross-sections merit critical study.

Ellis and Bahcall first calculated the cross-section for the reaction 
\begin{equation}
\nu _{e}+d\rightarrow e^{-}+p+p
\end{equation}
in 1968\cite{ebcc}, to aid in an early attempt to build a deuterium solar
neutrino detector. Later work by other authors~\cite{adnc,HCLee} focused on using the
reaction 
\begin{equation}
\nu _{e}+d\rightarrow \nu _{e}+n+p
\end{equation}
as a probe for neutral currents and a test of various electroweak models of
the time, in particular the Glashow-Salam-Weinberg model~\cite{GSW}.

The SNO proposal\cite{Ardsma,SNO} for a large scale heavy-water solar
neutrino detector led to a resurgence of interest in both processes, and to the
need for precise theoretical calculations of the cross-sections up through
supernovae neutrino energies of order 20~MeV, and beyond. While there have
been a number of calculations of progressive complexity~\cite
{ebcc,adnc,HCLee,BKN,TKK,dksno}, there are two definitive calculations of
these cross-sections which include modern nucleon-nucleon potentials and
meson-exchange current effects, though the approaches are quite different 
\cite{YHH,KN}. The differences between these calculations are quite small,
and it would appear that the results have converged.

However, there are no definitive experimental tests of these calculations,
yet SNO must rely on them in order to extract information on the solar
neutrino flux. As such, it is important that 
the physics behind these calculations is understood thoroughly, along
with any underlying theoretical and systematic uncertainties.

Our approach to the problem of neutrino-deuteron scattering is to take 
advantage of recent developments in the use of
low-energy effective field theory (EFT)
\cite{Weinberg1,Bira,Friara,Parka,cohena,Sa96,GPLa,LM,DR,KSW,Kolck,KSW2,CGSSpol,Ccompt,SSpv,KSSWpv,SSWst,MehStew,GruSho,Geg,Int,SteFurn,CohHan,Parkeft,Kong,Epel,MSWsu,PBB,threebod,PMRsupp,CRS,BMPV,KSte}.
The power counting scheme of Kaplan, Savage, and Wise~\cite{KSW} allows
for a concise and systematic analysis, order by order in perturbation theory,
and has been used
to study dynamical processes involving the
deuteron, including electromagnetic form-factors and moments\cite{KSW2},
Compton scattering\cite{CGSSpol,Ccompt}, $np\rightarrow d\gamma $\cite
{KSSWpv,SSWst}, and $pp\rightarrow de^{+}\nu _{e}$, where
electromagnetic effects are important\cite{Kong}. 
It is
straightforward to analyze neutrino-deuteron scattering in this scheme.
First, we will look at elastic $\nu $-$d$ scattering because of its
sensitivity to the strangeness property of the deuteron. Then, we will
proceed to two processes of interest to SNO, the neutral current (NC)
reaction $\nu +d\rightarrow \nu +n+p$ and the charged current (CC) reaction $%
\bar{\nu}+d\rightarrow e^{+}+n+n$. The normal CC
reaction, $\nu _{e}+d\rightarrow e^{-}+p+p$, is not considered here
because of the additional complication of
electromagnetic effects in the two nucleon final state.

\section{The Lagrangian}

\subsection{Effective Field Theory}

The lagrangian for an effective field theory involving nucleons and mesons
can be described via 
\begin{equation}
{\cal L}={\cal L}_{0}+{\cal L}_{1}+{\cal L}_{2}+\cdots \quad ,
\end{equation}
where ${\cal L}_{n}$ contains operators involving $n$ nucleons. Neglecting,
for the moment, the weak-interaction couplings 
\begin{equation}
{\cal L}_{0}={%
{\displaystyle{f^{2} \over 8}}%
}{\rm Tr}D_{\mu }\Sigma D^{\mu }\Sigma ^{\dagger }+{%
{\displaystyle{f^{2} \over 4}}%
}\lambda {\rm Tr}\,m_{q}(\Sigma +\Sigma ^{\dagger })+\cdots \quad ,
\end{equation}
where $\Sigma $ is the conventional unitary representation of the pion
fields in SU(2) 
\begin{equation}
\Sigma =\exp \left( 
{\displaystyle{2i\Pi  \over f}}%
\right) \quad ,\quad \Pi =\left( 
\begin{array}{cc}
\pi ^{0}/\sqrt{2} & \pi ^{+} \\ 
\pi ^{-} & -\pi ^{0}/\sqrt{2}
\end{array}
\right) \quad ,
\end{equation}
$f=132$ MeV is the pion decay constant, and the trace of the quark mass
matrix $m_{q}$ is related to the pion mass $m_{\pi }$ through $m_{\pi
}^{2}=\lambda (m_{u}+m_{d})$. The single nucleon lagrangian is 
\begin{equation}
{\cal L}_{1}=N^{\dagger }\bigg(iD_{0}+{\frac{{\bf D}^{2}}{2M_{N}}}\bigg)N-i{%
\frac{g_{A}}{2}}N^{\dagger }{\sigma }\cdot (\xi {\bf D}\xi ^{\dagger }-\xi
^{\dagger }{\bf D}\xi )N+\cdots \quad ,
\end{equation}
where the $\sigma $ Pauli matrix acts on nucleon spin space and $\xi =\exp
(i\Pi /f)$. The nucleon axial coupling is $g_{A}=1.26.$

The two nucleon lagrangian needed for next-to-leading order (NLO)
calculations is

\begin{eqnarray}
{\cal L}_{2} &=&-(C_{0}^{(\,^{3}S_{1})}+D_{2}^{(\,^{3}S_{1})}\lambda {\rm Tr}%
\,m_{q})(N^{T}P_{i}N)^{\dagger }(N^{T}P_{i}N)  \nonumber \\
&&+{\frac{C_{2}^{(\,^{3}S_{1})}}{8}}\left[ (N^{T}P_{i}N)^{\dagger }(N^{T}(%
\overleftarrow{{\bf D}}^{2}P_{i}-2\overleftarrow{{\bf D}}\cdot P_{i}%
\overrightarrow{{\bf D}}+P_{i}\overrightarrow{{\bf D}}^{2})N)+h.c.\right] 
\nonumber \\
&&-(C_{0}^{(\,\,^{1}S_{0})}+D_{2}^{(\,\,^{1}S_{0})}\lambda {\rm Tr}%
\,m_{q})(N^{T}\overline{P}_{i}N)^{\dagger }(N^{T}\overline{P}_{i}N) 
\nonumber \\
&&+{\frac{C_{2}^{(\,\,^{1}S_{0})}}{8}}\left[ (N^{T}\overline{P_{i}}%
N)^{\dagger }(N^{T}(\overleftarrow{{\bf D}}^{2}\overline{P}_{i}-2%
\overleftarrow{{\bf D}}\cdot \overline{P}_{i}\overrightarrow{{\bf D}}+%
\overline{P}_{i}\overrightarrow{{\bf D}}^{2})N)+h.c.\right] \quad ,
\end{eqnarray}
where $P_{i}$ and $\overline{P}_{i}$ are spin-isospin projectors for the $%
^{3}S_{1}$ channel and the $^{1}S_{0}$ channel, respectively, with definition
and normalization
\begin{eqnarray}
P_{i} &\equiv &\frac{1}{8}\sigma _{2}\sigma _{i}\tau _{2}\quad ,\quad \text{%
Tr}P_{i}^{\dagger }P_{j}=\frac{1}{2}\delta _{ij}\quad ,  \nonumber \\
\overline{P}_{i} &\equiv &\frac{1}{8}\sigma _{2}\tau _{2}\tau _{i}\quad
,\quad \text{Tr}\overline{P}_{i}^{\dagger }\overline{P}_{j}=\frac{1}{2}%
\delta _{ij}\quad ,
\end{eqnarray}
where the $\tau $ matrices act on isospin indices. The strong coupling
constants $C_{0}$, $C_{2}$ and $D_{2}$ have renormalization scale ($\mu $)
dependence. The details of the fitting procedure for these parameters, along
with their values in each channel, can be found in the Appendix.

\subsection{Weak Interactions}

The effective lagrangians for charged and neutral current weak interactions,
in terms of neutrino, nucleon, and meson fields, are given by 
\begin{equation}
{\cal L}_{{}}^{CC}\ =-{\displaystyle{\frac{G_{F}}{\sqrt{2}}}}l_{+}^{\mu
}J_{\mu }^{-}+h.c.\quad ,
\end{equation}
\begin{equation}
{\cal L}_{{}}^{NC}\ =-{\displaystyle{\frac{G_{F}}{\sqrt{2}}}}l_{Z}^{\mu
}J_{\mu }^{Z}\quad ,
\end{equation}
where the $l_{\mu }$ is the leptonic current and $J_{\mu }$ is the hadronic
current. For $\nu $-$d$ and $\overline{\nu }$-$d$ scattering, 
\begin{equation}
l_{+}^{\mu }=\overline{\nu }\gamma ^{\mu }(1-\gamma _{5})e\quad ,\quad
l_{Z}^{\mu }=\overline{\nu }\gamma ^{\mu }(1-\gamma _{5})\nu \quad .
\end{equation}
The hadronic currents can be decomposed into vector and axial-vector
contributions 
\begin{eqnarray}
J_{\mu }^{-} &=&V_{\mu }^{-}-A_{\mu }^{-}=(V_{\mu }^{1}-A_{\mu
}^{1})-i(V_{\mu }^{2}-A_{\mu }^{2})\quad ,  \nonumber \\
J_{\mu }^{Z} &=&-2\sin ^{2}\theta _{W}V_{\mu }^{S}+(1-2\sin ^{2}\theta
_{W})V_{\mu }^{3}-A_{\mu }^{S}-A_{\mu }^{3}\quad ,
\end{eqnarray}
where the superscripts represent isovector components (with $S$ representing
isoscalar terms) and, later, the currents will be labeled by the number of
nucleons involved.

In a next-to-leading order (NLO) calculation, the electron mass $m_{e}$
contributions to the matrix elements are counted as higher order,
along with contributions to the current proportional to the momentum
transfer $q_{\mu }$ since $q_{\mu }l^{\mu }=0$ up to NLO.
Weak-couplings to pion
fields are also higher order and are neglected here. The non-relativistic
one-body isoscalar currents are given by
\begin{eqnarray}
V_{0}^{S^{(1)}} &=&{\frac{1}{2}}N^{\dagger }N\quad ,  \nonumber \\
A_{0}^{S^{(1)}} &=&-{\frac{i}{2}}\Delta sN^{\dagger }{\frac{{\bf \sigma }%
\cdot (\overleftarrow{{\bf \nabla }}-\overrightarrow{{\bf \nabla }})}{2M_{N}}%
}N\quad ,  \nonumber \\
V_{k}^{S^{(1)}} &=&{-(}\kappa ^{(0)}+\frac{\mu _{s}}{4\sin ^{2}\theta _{W}}%
)N^{\dagger }\epsilon _{kij}{\frac{\sigma _{i}(\overleftarrow{\nabla }_{j}+%
\overrightarrow{\nabla }_{j})}{2M_{N}}}N\quad ,  \nonumber \\
A_{k}^{S^{(1)}} &=&-{\frac{1}{2}}\Delta sN^{\dagger }\sigma _{k}N\quad .
\label{current1}
\end{eqnarray}
Similarly, the isovector currents are given by
\begin{eqnarray}
V_{0}^{a^{(1)}} &=&N^{\dagger }{\frac{\tau ^{a}}{2}}N\quad ,  \nonumber \\
A_{0}^{a^{(1)}} &=&ig_{A}N^{\dagger }{\frac{\tau ^{a}}{2}}{\frac{{\bf \sigma 
}\cdot (\overleftarrow{{\bf \nabla }}-\overrightarrow{{\bf \nabla }})}{2M_{N}%
}}N\quad ,  \nonumber \\
V_{k}^{a^{(1)}} &=&-2\kappa ^{(1)}N^{\dagger }{\frac{\tau ^{a}}{2}}\epsilon
_{kij}{\frac{\sigma _{i}(\overleftarrow{\nabla }_{j}+\overrightarrow{\nabla }%
_{j})}{2M_{N}}}N\quad ,  \nonumber \\
A_{k}^{a^{(1)}} &=&g_{A}N^{\dagger }{\frac{\tau ^{a}}{2}}\sigma _{k}N\quad .
\label{current2}
\end{eqnarray}
There are vector currents from
magnetic moment terms, with
$\kappa ^{(0)}=\frac{1}{2}(\kappa _{p}+\kappa _{n})$ and 
$\kappa^{(1)}=\frac{1}{2}(\kappa _{p}-\kappa _{n})$ being the isoscalar and isovector
nucleon magnetic moments in nuclear magnetons, with 
\begin{equation}
\kappa _{p}=2.79285\quad ,\quad \kappa _{n}=-1.91304\quad .
\end{equation}
The isoscalar weak currents are also sensitive to strangeness matrix elements 
between nucleon states.
$\Delta s$ measures the strange quark contribution to
the proton spin 
\begin{equation}
2S_{\mu }\Delta s\equiv \left\langle p\left| \overline{s}\gamma _{\mu
}\gamma _{5}s\right| p\right\rangle \quad  \label{ds}
\end{equation}
with the value determined by Savage and Walden 
\begin{equation}
\Delta s=-0.17\pm 0.17
\end{equation}
from an analysis of lepton scattering data \cite{deltas1,deltas2} including
SU(3) symmetry breaking effects \cite{deltas}.
$S_{\mu }$ is the covariant spin vector. $\mu _{s}$ is
the strange magnetic moment of the proton 
\begin{gather}
\left\langle p\left| \overline{s}\gamma _{\mu }s\right| p\right\rangle =%
\overline{u}_{p}(G_{E}^{s}(q^{2})+G_{M}^{s}(q^{2})\frac{i\sigma _{\mu \nu
}q^{\nu }}{2M_{N}})u_{p}\quad ,  \nonumber \\
G_{E}^{s}(0)=0\quad ,\quad \mu _{s}\equiv G_{M}^{s}(0)\quad .  \label{mus}
\end{gather}
The value of $\mu _{s}$ has a large uncertainty. The SAMPLE experiment \cite
{SAMPLE} has measured 
\begin{equation}
G_{M}^{s}(-0.1{\rm GeV}^{2})=0.23\pm 0.37\pm 0.15\pm 0.19\text{ }{\rm n.m.}%
\quad ,
\end{equation}
However, theoretical predictions for $\mu _{s}$ range from $-0.8$ n.m. to 
$0.18$
n.m., as summarized in ref.~\cite{strangetheory}. 

Finally, there are two two-body axial currents relevant to scattering at
NLO: 
\begin{eqnarray}
A_{k}^{S^{(2)}} &=&-2i\varepsilon _{ijk}L_{2,A}(N^{T}P_{i}N)^{\dagger
}(N^{T}P_{j}N)\quad ,  \nonumber \\
A_{k}^{a^{(2)}} &=&L_{1,A}\left( N^{T}P_{k}N\right) ^{\dagger }\left( N^{T}%
\overline{P}_{a}N\right) +h.c.\quad .  \label{2body}
\end{eqnarray}
In addition,  there is a two-body vector current that contributes to the
NLO strange magnetic form factor 
\begin{equation}
V_{k}^{S^{(2)}}=2iL_{2}^{s}(N^{T}P_{i}N)^{\dagger }(\overleftarrow{\nabla }%
_{i}+\overrightarrow{\nabla }_{i})(N^{T}P_{k}N)+h.c.\quad .  \label{2bodyx}
\end{equation}

\section{Neutrino-Deuteron and Antineutrino-Deuteron Elastic Scattering}

\subsection{Deuteron Form Factors}

$\nu d\rightarrow \nu d$ and $\overline{\nu }d\rightarrow \overline{\nu }d$
elastic scattering processes are sensitive to
the strange quark properties of the deuteron. This can be easily seen by
writing the neutral current in terms of quark degrees of freedom.
\begin{equation}
J_{\mu }^{Z}=\frac{1}{2}\left[ \overline{u}\gamma _{\mu }(1-\gamma _{5})u-%
\overline{d}\gamma _{\mu }(1-\gamma _{5})d-\overline{s}\gamma _{\mu
}(1-\gamma _{5})s\right] -2\sin ^{2}\vartheta _{W}J_{\mu }^{em}\quad ,
\end{equation}
then taking the matrix element between deuteron states. The fact that the
deuteron is isoscalar means that the matrix element depends only on the
electromagnetic and strangeness properties of the deuteron, 
\begin{eqnarray}
\langle \ d\left| J_{\mu }^{Z}\right| d\rangle &=&\left\langle d\left|
-2\sin ^{2}\vartheta _{W}J_{\mu }^{em}-\frac{1}{2}\overline{s}\gamma _{\mu
}s\right| d\right\rangle +\left\langle \ d\left| \frac{1}{2}\overline{s}%
\gamma _{\mu }\gamma _{5}s\right| d\right\rangle  \nonumber \\
&\equiv &\quad \left\langle \ d\left| V_{\mu }^{Z}\right| d\right\rangle
-\left\langle \ d\left| A_{\mu }^{Z}\right| d\right\rangle \quad .
\label{NCcurrent}
\end{eqnarray}
Since the deuteron's electromagnetic properties are well known, $\nu d$ and $%
\overline{\nu }d$ elastic scattering processes are, in theory, ideal for studying
the strangeness of the deuteron and might also provide valuable
information on strangeness in the nucleon.

To begin, we must define the NC vector and axial form factors of the
deuteron, then calculate the strange form factors in EFT up to NLO. Using $%
\left| {\bf p,}i\right\rangle $ to represent a deuteron with momentum ${\bf p%
}$ and polarization state $i$, the vector form factors of the neutral
weak current can be defined in the way similar to that of the
electromagnetic form factors

\begin{equation}
\langle \ {\bf p}^{^{\prime }},j\left| V_{0}^{Z}\right| {\bf p,}i\rangle =%
\left[ F_{C}^{Z}({\bf q}^{2})\delta _{ij}+%
{\displaystyle{1 \over 2M_{d}^{2}}}%
F_{Q}^{Z}({\bf q}^{2})({\bf q}_{i}{\bf q}_{j}-%
{\displaystyle{1 \over 3}}%
{\bf q}^{2}\delta _{ij})\right] \quad ,
\end{equation}
\begin{eqnarray}
\left\langle \ {\bf p}^{^{\prime }},j\left| V_{k}^{Z}\right| {\bf p,}%
i\right\rangle &=&%
{\displaystyle{1 \over 2M_{d}}}%
[F_{C}^{Z}({\bf q}^{2})\delta _{ij}({\bf p}+{\bf p}^{^{\prime
}})_{k}+F_{M}^{Z}({\bf q}^{2})(\delta _{i}^{k}{\bf q}_{j}-\delta _{j}^{k}%
{\bf q}_{i})  \nonumber \\
&&+%
{\displaystyle{1 \over 2M_{d}^{2}}}%
F_{Q}^{Z}({\bf q}^{2})({\bf q}_{i}{\bf q}_{j}-\frac{1}{3}{\bf q}^{2}\delta
_{ij})({\bf p}+{\bf p}^{^{\prime }})_{k}]\quad ,
\end{eqnarray}
where $M_{d}$ is the deuteron mass and ${\bf q=p}^{^{\prime }}-{\bf p}$. The
neutral weak charge, magnetic and quadrupole form factors $F_{C}^{Z}$, $%
F_{M}^{Z}$ and $F_{Q}^{Z}$ are linear combinations of the electromagnetic
charge, magnetic and quadrupole form factors $F_{C}$, $F_{M}$ and $F_{Q}$
and the strange charge, magnetic and quadrupole form factors $F_{C}^{s}$, $%
F_{M}^{s}$ and $F_{Q}^{s}$. 
\begin{eqnarray}
F_{C}^{Z} &=&-2\sin ^{2}\vartheta _{W}F_{C}-\frac{1}{2}F_{C}^{s}\quad , 
\nonumber \\
F_{M}^{Z} &=&-2\sin ^{2}\vartheta _{W}F_{M}-\frac{1}{2}F_{M}^{s}\quad , 
\nonumber \\
F_{Q}^{Z} &=&-2\sin ^{2}\vartheta _{W}F_{Q}-\frac{1}{2}F_{Q}^{s}\quad .
\end{eqnarray}
The electromagnetic form factors of the deuteron are well
known experimentally, and have been computed in \cite{KSW2} using effective field theory.
The form factor $F_{C}^{s}=0$ at ${\bf q}^{2}=0$, because there are no net
strange quarks in the deuteron. This means that $F_{C}^{s}$ , together with $%
F_{Q}^{s}$ do not contribute to the cross-section until next-next-to-leading
order (NNLO) and will not be considered in this paper.

The axial form factor 
\begin{eqnarray}
\left\langle \ {\bf p}^{^{\prime }},j\left| A_{k}^{Z}\right| {\bf p,}%
i\right\rangle &=&i\varepsilon _{jki}F_{A}^{Z}({\bf q}^{2})\quad ,  \nonumber
\\
\left\langle \ {\bf p}^{^{\prime }},j\left| A_{0}^{Z}\right| {\bf p,}%
i\right\rangle &=&i\varepsilon _{jki}F_{A}^{Z}({\bf q}^{2})%
{\displaystyle{({\bf p}+{\bf p}^{^{\prime }})_{k} \over 2M_{d}}}%
\quad ,  \label{axial}
\end{eqnarray}
arises directly from strange matrix elements only, as has been noted from 
eq.(\ref{NCcurrent}) already. 
\begin{equation}
F_{A}^{Z}=\ F_{A}^{s}\quad .
\end{equation}
Note again that terms proportional to $q_{\mu }$ have been dropped in eq.(\ref{axial}%
) since we treat the neutrino as massless.

The effective field theory result for the strange form factors $%
F_{A}^{s}$ and $F_{M}^{s}$ up to NLO can be written as a
series expansion in powers of\ $Q$, where $Q$ is the small expansion
parameter in the momentum and chiral expansion 
\begin{equation}
F^{s}=F^{s(0)}+F^{s(1)}+\cdots \quad .
\end{equation}
The superscripts denote the power of $Q$ in the expansion. The LO strange
axial form factor depends on the strange axial moment of the nucleon $\Delta
s$ (defined in eq.(\ref{ds})) and the LO electric form factor $F_{C}^{(0)}$ 
\begin{equation}
F_{A}^{s(0)}({\bf q}^{2})=-\Delta sF_{C}^{(0)}({\bf q}^{2})\ \quad ,
\end{equation}
where 
\begin{equation}
F_{C}^{(0)}({\bf q}^{2})=\frac{4\gamma }{\left| {\bf q}\right| }\tan ^{-1}%
{\left| {\bf q}\right|  \overwithdelims() 4\gamma }%
\quad .  \label{fc0}
\end{equation}
$\gamma =\sqrt{M_{N}B}$, and $B=2.2245$ MeV is the deuteron binding energy.

The NLO strange axial form factor depends on the NLO electric form factor $%
F_{C}^{(1)}$, the electric quadrupole moment of the first non-vanishing
order $F_{Q}^{(-1)}$, and a two-body operator with coefficient $L_{2,A}$
from eq.(\ref{2body}). 
\begin{equation}
F_{A}^{s(1)}({\bf q}^{2})=-\Delta s(F_{C}^{(1)}({\bf q}^{2})+\frac{{\bf q}%
^{2}}{12M_{d}^{2}}F_{Q}^{(-1)}({\bf q}^{2}))+L_{2,A}(\mu )\frac{\gamma (\mu
-\gamma )^{2}}{\pi }\quad .
\end{equation}
$L_{2,A}(\mu )$ itself depends on $\mu$ but the form factor is 
$\mu$-independent. The
expressions for $F_{C}^{(1)}$ and $F_{Q}^{(-1)}$ can be found in \cite
{KSW2}.

The strange magnetic form factor is computed as 
\begin{align}
F_{M}^{s(0)}({\bf q}^{2})& =4\mu _{s}F_{C}^{(0)}({\bf q}^{2})\quad , 
\nonumber \\
\ F_{M}^{s(1)}({\bf q}^{2})& =4\mu _{s}(F_{C}^{(1)}({\bf q}^{2})+\frac{{\bf q%
}^{2}}{12M_{d}^{2}}F_{Q}^{(-1)}({\bf q}^{2}))+4L_{2}^{s}(\mu )\frac{%
M_{N}\gamma (\mu -\gamma )^{2}}{\pi }\quad ,
\end{align}
where $\mu _{s}$ is the strange magnetic moment of the nucleon defined in
eq.(\ref{mus}) and $L_{2}^{s}$ is the coefficient of the two-body counter
term defined in eq.(\ref{2bodyx}).

\subsection{ Cross-Section}

The $\nu (\overline{\nu })d$ elastic scattering cross-section up to NLO is 
\begin{eqnarray}
{\displaystyle{d\sigma  \over d\Omega }}%
_{\nu (\overline{\nu })d}=%
{\displaystyle{G_{F}^{2}{\omega }^{2} \over 2\pi ^{2}}}%
&\bigg\{& \cos ^{2}%
{\displaystyle{{\theta } \over 2}}%
(4\sin ^{4}\vartheta _{W}\ F_{C}^{2}+\ 
{\displaystyle{2 \over 3}}%
(F_{A}^{s})^{2})\nonumber\\
&&+2\sin ^{2}%
{\displaystyle{{\theta } \over 2}}%
\left[ 
{\displaystyle{2 \over 3}}%
(F_{A}^{s})^{2}\mp 
{\displaystyle{2{\omega } \over 3M_{d}}}%
(4\sin ^{2}\vartheta _{W}F_{M}+F_{M}^{s})F_{A}^{s}\right] \bigg\} \quad ,
\label{elastic}
\end{eqnarray}
where ${\omega }$ is the incident neutrino (antineutrino) energy, and ${\theta }$ is the 
scattering angle between the incident and outgoing lepton
directions. The negative sign of the magnetic and axial interference term
corresponds to $\nu d$ scattering and the positive sign to $\overline{\nu }d$. 
Note that in addition to the $F_{M}^{s}$ dependence which is not
considered in \cite{FHPY}, we also disagree on the sign of the interference
term. In our expression, $\sigma _{\nu d}<\sigma _{\overline{\nu }d}$ when $%
F_{M}^{s}=0$. This can be confirmed, later, when we consider the
elastic limit of the inelastic scattering process.

In reference\cite{FHPY}, the motivation was to measure $\Delta s$ through
the ratio 
\begin{equation}
R_{el}={\frac{d\sigma (\nu d\rightarrow \nu d)-d\sigma (\bar{\nu}d\rightarrow 
\bar{\nu}d)}{d\sigma (\nu d\rightarrow \nu d)+d\sigma (\bar{\nu}d\rightarrow 
\bar{\nu}d)}}\text{\quad .}
\end{equation}
At LO, this ratio depends on $\Delta s$ and $\mu _{s}$
while at NLO it also depends on $L_{2,A}$ and $L_{2}^{s}$. Given that the
values of $\Delta s$ and $\mu _{s}$ should become known more precisely from
single-nucleon experiments, precision
measurements of the ratio $R_{el}$ could tell us more about the intrinsic
strangeness of the deuteron.

\section{Neutrino-Deuteron and Antineutrino Deuteron Neutral Current Inelastic Scattering}

\subsection{Neutral Current Structure Factors}

For neutrino scattering from a hadronic target,
the differential cross-section can be written in terms of leptonic and
hadronic tensors $l_{\mu \nu }$ and $W_{\mu \nu }$ as 
\begin{equation}
{\displaystyle{d^{2}\sigma  \over d\omega ^{^{\prime }}d\Omega }}%
=%
{\displaystyle{G_{F}^{2}|{\bf k}^{\prime }| \over 32\pi ^{2}|{\bf k}| }}%
\ l^{\mu \nu }W_{\mu \nu }\quad ,
\label{crosssec}
\end{equation}
where ${\bf k} ({\bf k}^\prime)$  represents the
initial (final) lepton three-momentum. The leptonic tensor is given by 
\begin{equation}
\ l^{\mu \nu }=8(k^{\mu }k^{\nu \prime }+k^{\nu }k^{\mu \prime }-k\cdot
k^{\prime }g^{\mu \nu }+i\varepsilon ^{\mu \nu \rho \sigma }k_{\rho
}k_{\sigma }^{^{\prime }})\quad .  \label{lepten}
\end{equation}
The hadronic tensor can be defined as the imaginary part of the forward 
matrix element
of the
time-ordered product of two weak current operators. 
It can be parameterized by
six different structure functions 
\begin{eqnarray}
W_{\mu \nu } &=&%
{\displaystyle{1 \over \pi }}%
\text{Im}\left[ 
\displaystyle\int %
d^{4}xe^{iqx}T\left\langle d(P)\left| J_{\mu }^{Z^{\dagger }}(x)J_{\nu
}^{Z}(0)\right| d(P)\right\rangle \right]  \nonumber \\
&=&-W_{1}g_{\mu \nu }+W_{2}%
{\displaystyle{P_{\mu }P_{\nu } \over M_{d}^{2}}}%
-iW_{3}\varepsilon _{\mu \nu \alpha \beta }%
{\displaystyle{P^{\alpha }q^{\beta } \over M_{d}^{2}}}%
\nonumber \\
&&+W_{4}%
{\displaystyle{q_{\mu }q_{\nu } \over M_{d}^{2}}}%
+W_{5}%
{\displaystyle{(P_{\mu }q_{\nu }+q_{\mu }P_{\nu }) \over M_{d}^{2}}}%
+iW_{6}%
{\displaystyle{(P_{\mu }q_{\nu }-q_{\mu }P_{\nu }) \over M_{d}^{2}}}%
\quad ,
\end{eqnarray}
where the momentum transfer $q_{\mu }=k_{\mu }-k_{\mu }^{\prime }$ (the difference
between the incident and outgoing lepton four-momenta), and $P_{\mu}$ is
the deuteron four-momentum. 
$W_{4},\ W_{5}$ and $W_{6}$ do not contribute to the
differential cross-section because $q_{u}l^{\mu \nu }=0$. 

For the reaction
\begin{equation}
\nu +d\rightarrow \nu +n+p,
\end{equation}
the differential cross-section in the lab frame
(deuteron rest frame) simplifies to
\begin{equation}
{\displaystyle{d^{2}\sigma  \over d\omega ^{^{\prime }}d\Omega }}%
=%
{\displaystyle{G_{F}^{2}\omega ^{\prime} |{\bf k}^\prime|\over 2\pi ^{2}}}%
\left[ 2W_{1}\sin ^{2}%
{\displaystyle{\theta  \over 2}}%
+W_{2}\cos ^{2}%
{\displaystyle{\theta  \over 2}}%
-2%
{\displaystyle{(\omega +\omega ^{\prime }) \over M_{d}}}%
W_{3}\sin ^{2}%
{\displaystyle{\theta  \over 2}}%
\right] \quad ,  \label{dsig}
\end{equation}
where $\theta $ is the angle between {\bf k} and {\bf k}$^{^{\prime }}$, $\omega^\prime$
is the final lepton energy,
and we have used the relation 
\begin{equation}
q^{2}=-4\omega \omega ^{\prime }\sin ^{2}%
{\displaystyle{\theta  \over 2}}%
\quad . \label{eqdiffcs}
\end{equation}
For $\overline{\nu }d\ \rightarrow \overline{\nu }np$ scattering, the last
terms on the right hand sides of eq.(\ref{lepten}) and (\ref{dsig}) change
sign.

The phase space boundaries for this reaction through the angular bound
\begin{equation}  \label{cosnc}
\text{Max}\left[ -1,1-%
{\displaystyle{4M_N(\nu -B)-\nu ^2 \over 2\omega \omega ^{\prime }}}%
\right] \leq \cos \theta \leq 1\quad ,
\end{equation}
and the bound on outgoing lepton energy
\begin{equation}
0\leq \omega ^{\prime }\leq \omega -2(M_N-\sqrt{M_N^2-\gamma ^2})\quad ,
\end{equation}
where $\nu =\omega -\omega ^{\prime }$ is the energy transfer ($q_0$).

\subsection{Leading order}

%
\begin{figure}[!t] 
\centerline{{\epsfxsize=5in \epsfbox{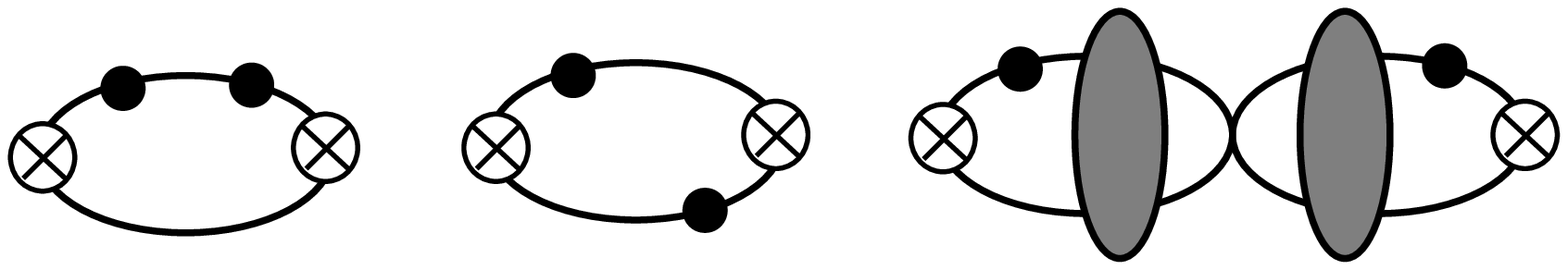}}}
\vskip 0.1in
\centerline{ {\epsfxsize=5in \epsfbox{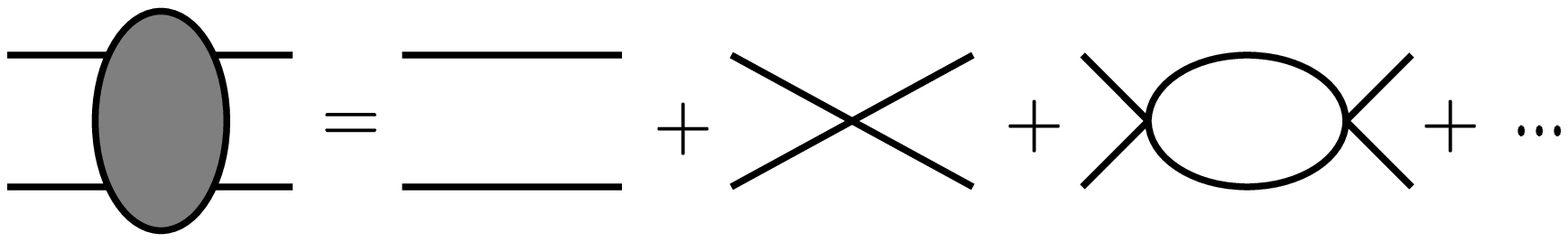}}}
\noindent

\caption{{\it Leading order contributions to the $\protect\nu $-$d$
scattering hadronic tensor. The gray blobs arise from no insertion plus any
numbers of insertions of four nucleon operators $C_{0}^{(^{3}S_{1})}$ or $%
C_{0}^{(^{1}S_{0})}$, as appropriate. The crossed circles denote operators that create or
annihilate two nucleons with the quantum numbers of the deuteron. The solid
lines are nucleons. The solid circles correspond to insertions of leading weak
current operators. }}
\label{figLO}
\end{figure}

The leading order contributions to the hadronic tensor $W_{\mu\nu}$ are the diagrams shown in
Fig.~\ref{figLO}.  The non-zero components of $W_{\mu\nu}$ are given by
\begin{eqnarray}
W_{00}^{LO}
&=&2(C_{V}^{(0)^{2}}+C_{V}^{(1)^{2}})F_{1}+(C_{V}^{(0)^{2}}-C_{V}^{(1)^{2}})F_{2}+4C_{V}^{(0)^{2}}F_{3}\quad ,
\nonumber \\
W_{ij}^{LO} &=&\delta _{ij}\left( 2(C_{A}^{(0)^{2}}+C_{A}^{(1)^{2}})F_{1}+%
{\displaystyle{1 \over 3}}%
(C_{A}^{(0)^{2}}-C_{A}^{(1)^{2}})F_{2}+\frac{8}{3}C_{A}^{(0)^{2}}F_{3}+\frac{%
4}{3}C_{A}^{(1)^{2}}F_{4}\right) \   \nonumber \\
&\equiv &\delta _{ij}W_{ii}^{LO}\quad ,  \label{w00LO}
\end{eqnarray}
with no summation implied over indexes of $W_{ii}^{LO}$.
The coefficients arise from the specific forms of the single-nucleon vector and axial
currents, and are given by
\begin{equation}
\begin{array}[!t]{ll}
C_{V}^{(0)}=-\sin ^{2}\vartheta _{W}\quad & ,\quad C_{V}^{(1)}=%
{\displaystyle{1 \over 2}}%
(1-2\sin ^{2}\vartheta _{W})\quad , \\ 
C_{A}^{(0)}=-%
{\displaystyle{1 \over 2}}%
\Delta s\quad & ,\quad C_{A}^{(1)}=%
{\displaystyle{1 \over 2}}%
g_{A}\quad .
\end{array}
\end{equation}

The functions $F_{a}$ are the individual contributions at LO from the
diagrams of Fig.~\ref{figLO}, with $F_{3}$ and $F_{4}$ both associated with
the last diagram in the first row:

\begin{eqnarray}
F_{1} &=&\text{Re}\left[ 
{\displaystyle{2M_{N}\gamma \ p \over \pi \left[ M_{N}^{2}\nu ^{2}-{\bf q}^{2}p^{2}\right] }}%
\right] \quad ,  \nonumber \\
F_{2} &=&\text{Re}\left[ 
{\displaystyle{4\ \gamma \  \over \pi \nu \left| {\bf q}\right| }}%
\tanh ^{-1}\left( 
{\displaystyle{\left| {\bf q}\right| p\  \over M_{N}\nu }}%
\right) \right] \quad ,  \nonumber \\
F_{3} &=&%
{\displaystyle{2\ \gamma \  \over M_{N}^{2}}}%
\text{Im}\left\{ \left[ 
{\displaystyle{M_{N}^{2}\  \over \pi \left| {\bf q}\right| }}%
\tan ^{-1}\left( 
{\displaystyle{\left| {\bf q}\right| \  \over 2(\gamma -ip)}}%
\right) \right] ^{2}A_{-1}^{(^{3}S_{1})}(p)\right\} \quad ,  \nonumber \\
F_{4} &=&F_3(^{3}S_{1}\rightarrow \ ^{1}S_{0})\ \quad .
\label{LOfun}
\end{eqnarray}
The magnitude of the relative momentum between the final state proton and
neutron is $p$, with
\begin{equation}
p=\sqrt{M_{N}\nu -\gamma ^{2}-\frac{{\bf q}^{2}}{4}+i\epsilon }\quad .
\end{equation}
The `rescattering' amplitudes appearing in $F_3$ and $F_4$ are the leading order
$NN$ scattering amplitudes in each channel, given by 
\begin{equation}
A_{-1}^{(^{3}S_{1})}(p)=%
{\displaystyle{-C_{0}^{(^{3}S_{1})} \over 1+%
{\displaystyle{M_{N}C_{0}^{(^{3}S_{1})}\  \over 4\pi }}\left( \mu +ip\right) }}%
\quad ,
\end{equation}
and similarly for the $^1S_0$ channel.
After matching onto the effective range expansion as performed in the Appendix,
\begin{equation}
A_{-1}^{(^{1}S_{0})}(p)=%
-{\displaystyle{4\pi  \over M_{N}}}%
{\displaystyle{1 \over %
{\displaystyle{1 \over a_{{}}^{(^{1}S_{0})}}}+ip}}%
\quad ,
\end{equation}
and
\begin{equation}
A_{-1}^{(^{3}S_{1})}(p)=%
-{\displaystyle{4\pi  \over M_{N}}}%
{\displaystyle{1 \over \gamma +ip}}%
\quad ,
\end{equation}
where $a^{(^{1}S_{0})}$ is the $^{1}S_{0}$ channel scattering length.

From these expressions for the
components of the hadronic tensor, we find that the leading contributions to the
structure factors in eq.(\ref{dsig}) are
\begin{eqnarray}
W_{1}^{LO} &=&W_{ii}^{LO}\quad ,  \nonumber \\
W_{2}^{LO} &=&W_{00}^{LO}+W_{ii}^{LO}\quad ,  \nonumber \\
W_{3}^{LO} &=&0\quad .  \label{w1w2}
\end{eqnarray}

\subsubsection{Elastic Limit}

An important test of the inelastic calculation is that we can reproduce the
elastic scattering results in the limit 
\begin{equation}
\ \qquad p\rightarrow i\gamma +\epsilon \quad .
\end{equation}
It is straightforward to see that, in this limit, 
\begin{eqnarray}
F_{1} &=&0\quad ,  \nonumber \\
F_{2} &=&0\quad ,  \nonumber \\
F_{3} &=&F_{C}^{(0)^{2}}(\left| {\bf q}\right| )\delta (\nu -%
{\displaystyle{{\bf q}^{2} \over 4M_{N}}}%
)\quad ,  \nonumber \\
F_{4} &=&0\ \quad .
\end{eqnarray} 
This, combined with eq.(\ref{dsig},\ref{w00LO},\ref{w1w2}),
reproduces the LO contribution to eq.(\ref{elastic}). Further it can be shown that
our NLO result also recovers the elastic limit.

\subsubsection{Threshold Behavior}

\label{thresh}Another useful test of the inelastic calculation is to study
the threshold behavior and compare that with the results expected from
the effective range expansion. In the effective range expansion, it is well
known that the dominant contribution to the threshold hadronic matrix element is
the $^{3}S_{1}\rightarrow $ $^{1}S_{0}$ transition through the isovector
axial coupling. The $^{3}S_{1}\rightarrow \,^{3}S_{1}$ transition is
suppressed because amongst the NC spin-isospin operators {\bf 1}, 
$\tau ^{a}$, 
$\sigma ^{i}$ and $\tau ^{a}\sigma ^{i}$: i) the isovector operators don't
contribute (the transition is isoscalar); and ii) the matrix elements of the
isoscalar operators vanish in the zero recoil limit ($d$ and $np$ states are
orthogonal in the zero recoil limit).

Our results reproduce these features. In the threshold limit $\omega ^{^{\prime
}},p\rightarrow 0$, 
\begin{eqnarray}
W_{00}^{LO} &=&0\quad ,  \nonumber \\
W_{ii}^{LO} &=&\ \left( 
{\displaystyle{2g_{A}^{2}M_{N}p \over 3\pi \gamma ^{3}}}%
\right) (1-a^{(^{1}S_{0})\ }\gamma )^{2}\quad .
\end{eqnarray}
Another
consequence of the $^{3}S_{1}\rightarrow \,^{3}S_{1}$ suppression is that
our results will not be sensitive to isoscalar parameters $\Delta s,$ $\mu _{s}$
and $L_{2,A}$. This has the added advantage of reducing the number of free
parameters, as will be seen later.

Using eqs.(\ref{dsig},\ref{cosnc},\ref{w1w2}), the differential cross-section
with respect to the relative kinetic energy $E_{r}(=p^{2}/M_{N})$ between
final state nucleons is 
\begin{eqnarray}
{\displaystyle{d\sigma  \over dE_{r}}}%
&=&%
\displaystyle\int %
_{-1}^{1}d(\cos \theta )2\pi %
{\displaystyle{d^{2}\sigma  \over d\omega ^{^{\prime }}d\Omega }}%
\nonumber \\
&=&%
{\displaystyle{2G_{F}^{2}g_{A}^{2}(\omega -B)^{2}M_{N}p \over \pi ^{2}\gamma ^{3}}}%
(1-a^{(^{1}S_{0})\ }\gamma )^{2}\quad .
\end{eqnarray}
This reproduces the threshold behavior of the effective range expansion
result of \cite{HCLee}. Our result is also consistent, at both LO and NLO,
with the analytic expression of the $np\rightarrow d\gamma $ amplitude given
in \cite{SSWst}.

\subsection{Next to Leading Order}

At NLO we will decompose the hadronic tensor into five components 
\begin{equation}
W_{\mu \nu }^{NLO}=W_{\mu \nu }^{D_{2}}+W_{\mu \nu }^{L_{A}}+W_{\mu \nu
}^{C_{2}}+W_{\mu \nu }^{\ \pi }+W_{\mu \nu }^{\ M_{Z}}\quad ,
\end{equation}
each corresponding to an insertion of a NLO ${\cal L}_{2}$ operator, single
pion exchange or higher order weak couplings. Each of these five components
will be considered separately.

\subsubsection{D$_2$ Contributions}

\begin{figure}[!t]
\centerline{{\epsfxsize=5.0in \epsfbox{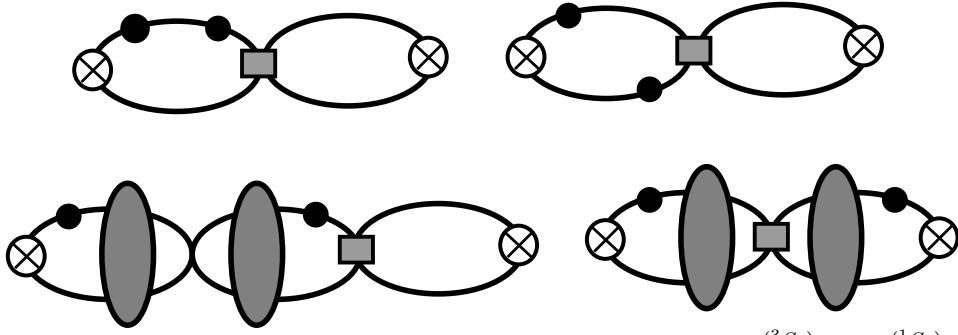}}}
\noindent
\caption{{\it Graphs from insertions of the operators with coefficients $%
D_{2}^{(^{3}S_{1})}$ or $D_{2}^{(^{1}S_{0})}$ (denoted by the gray squares)
that contribute to the $\protect\nu $-$d$ scattering hadronic tensor at NLO.
The gray blobs and other features are as defined in Fig.~\ref{figLO}.
}}
\label{figD2}
\end{figure}

The $D_{2}$ contributions to the hadronic tensor arise from the diagrams
shown in Fig.~\ref{figD2}. They can be expressed as simple replacements of the LO
results in eqs.~(\ref{w00LO},\ref{LOfun}),

%

\begin{equation}
W_{\mu \nu }^{D_{2}}=W_{\mu \nu }^{LO}(F_{1}\rightarrow 0,F_{2}\rightarrow
0,A_{-1}^{(^{3}S_{1})}(p)\rightarrow
A_{0,D_{2}}^{(^{3}S_{1})}(p),A_{-1}^{(^{1}S_{0})}(p)\rightarrow
A_{0,D_{2}}^{(^{1}S_{0})}(p))\quad ,
\end{equation}
where 
\begin{equation}
A_{0,D_{2}}^{(^{3}S_{1})}(p)\
=-D_{2}^{(^{3}S_{1})}m_{\pi }^{2}\left[ 
{\displaystyle{A_{-1}^{(^{3}S_{1})}(p) \over C_{0}^{(^{3}S_{1})}}}%
\right] ^{2}\quad ,
\end{equation}
and similarly for the $^1S_0$ channel.

\subsubsection{L$_{A}$ Contributions}

\begin{figure}[!t]  
\centerline{{\epsfxsize=5.0in \epsfbox{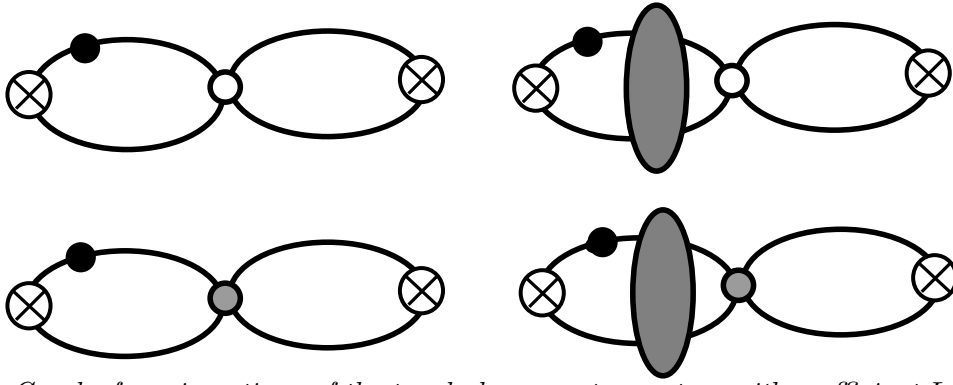}}}
\noindent

\caption{{\it Graphs from insertions of the two-body current operators with coefficient $%
L_{1,A}$ (denoted by the circles) and $L_{2,A}$ (gray circles) that
contribute to the $\protect\nu $-$d$ scattering hadronic tensor at NLO. 
The gray blobs and other features are as defined in Fig.~\ref{figLO}.
 }}
\label{figLa}
\end{figure}
%

Diagrams with one insertion of $L_{1,A}$ or $L_{2,A}$ contribute to the
hadronic tensor at NLO are shown in Fig.~\ref{figLa}. These contribute only
to the spatial piece of the hadronic tensor, and we find 
\begin{equation}
W_{ii}^{L_{A}}=-%
{\displaystyle{4M_{N}\gamma (\mu -\gamma ) \over 3\pi ^{2}\left| {\bf q}\right| }}%
\text{Im}\left\{ \tan ^{-1}\left( 
{\displaystyle{\left| {\bf q}\right| \  \over 2(\gamma -ip)}}%
\right) \left[ 
{\displaystyle{C_{A}^{(1)}L_{1,A} \over C_{0}^{(^{1}S_{0})}}}%
A_{-1}^{(^{1}S_{0})}(p)+%
{\displaystyle{4C_{A}^{(0)}L_{2,A} \over C_{0}^{(^{3}S_{1})}}}%
A_{-1}^{(^{3}S_{1})}(p)\right] \right\} \quad ,
\end{equation}
with all other $W_{\mu \nu }^{L_{A}}=0.$

Note that the term involving $L_{1,A}$ behaves like 
\begin{equation}
{\displaystyle{\ L_{1,A}(\mu -\gamma ) \over C_{0}^{(^{1}S_{0})}}}%
\quad ,
\end{equation}
which is {\em not }$\mu $--independent. The $\beta $ function of $L_{1,A}$
can be obtained by requiring the $NN\rightarrow NN\nu \overline{\nu },$ $%
^{3}S_{1}$ to $^{1}S_{0}$ transition amplitude to be $\mu $--independent, 
\begin{equation}
\mu 
{\displaystyle{d \over d\mu }}%
\left( 
{\displaystyle{L_{1,A}-C_{A}^{(1)}M_{N}(C_{2}^{(^{1}S_{0})}+C_{2}^{(^{3}S_{1})}) \over C_{0}^{(^{1}S_{0})}C_{0}^{(^{3}S_{1})}}}%
\right) =0\quad .  \label{l1beta}
\end{equation}

\subsubsection{C$_2$ Contributions}

\begin{figure}[!t]
\centerline{{\epsfxsize=5.0in \epsfbox{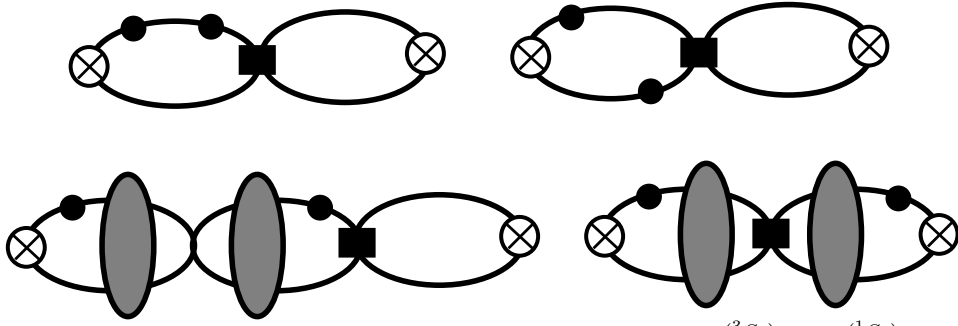}}}
\noindent

\caption{{\it Graphs from insertions of the operator with coefficient $%
C_{2}^{(^{3}S_{1})}$ or $C_{2}^{(^{1}S_{0})}$ (denoted by the solid squares)
that contribute to the $\protect\nu $-$d$ scattering hadronic tensor at NLO.
The gray blobs and other features are as defined in Fig.~\ref{figLO}.
}}
\label{figc2}
\end{figure}
%

 Fig.~\ref{figc2} shows the diagrams in which $C_{2}$ contributes to the hadronic
tensor at NLO.  Here we find contributions to both spatial and time-like components
of the hadronic tensor
\begin{eqnarray}
W_{00}^{C_{2}} &=&%
{\displaystyle{M_{N}\gamma C_{2}^{(^{3}S_{1})}(\mu -\gamma )^{2} \over 2\pi }}%
W_{00}^{LO}+4C_{V}^{(0)^{2}}F_{3}(A_{-1}^{(^{3}S_{1})}\rightarrow
A_{0,C_{2}}^{(^{3}S_{1})})  \nonumber \\
&&+%
{\displaystyle{8C_{V}^{(0)^{2}}C_{2}^{(^{3}S_{1})}M_{N}^{2}\gamma (\mu -\gamma ) \over \pi ^{2}\left| {\bf q}\right| \ C_{0}^{(^{3}S_{1})}}}%
\text{Im}\left[ A_{-1}^{(^{3}S_{1})}(p)\tan ^{-1}\left( 
{\displaystyle{\left| {\bf q}\right| \  \over 2(\gamma -ip)}}%
\right) \right] \quad ,
\end{eqnarray}
and 
\begin{eqnarray}
W_{ii}^{C_{2}} &=&%
{\displaystyle{M_{N}\gamma C_{2}^{(^{3}S_{1})}(\mu -\gamma )^{2} \over 2\pi }}%
W_{ii}^{LO}  \nonumber \\
&&+%
{\displaystyle{8 \over 3}}%
C_{A}^{(0)^{2}}F_{3}\left( A_{-1}^{(^{3}S_{1})}\rightarrow
A_{0,C_{2}}^{(^{3}S_{1})}\right) +%
{\displaystyle{4 \over 3}}%
C_{A}^{(1)^{2}}F_{4}\left( A_{-1}^{(^{1}S_{0})}\rightarrow
A_{0,C_{2}}^{(^{1}S_{0})}\right)  \nonumber \\
&&+%
{\displaystyle{16C_{A}^{(0)^{2}}C_{2}^{(^{3}S_{1})}M_{N}^{2}\gamma (\mu -\gamma ) \over 3\pi ^{2}\left| {\bf q}\right| \ C_{0}^{(^{3}S_{1})}}}%
\text{Im}\left[ A_{-1}^{(^{3}S_{1})}(p)\tan ^{-1}\left( 
{\displaystyle{\left| {\bf q}\right| \  \over 2(\gamma -ip)}}%
\right) \right]  \nonumber \\
&&+%
{\displaystyle{4C_{A}^{(1)^{2}}(C_{2}^{(^{1}S_{0})}+C_{2}^{(^{3}S_{1})})M_{N}^{2}\gamma (\mu -\gamma ) \over 3\pi ^{2}\left| {\bf q}\right| \ C_{0}^{(^{1}S_{0})}}}%
\text{Im}\left[ A_{-1}^{(^{1}S_{0})}(p)\tan ^{-1}\left( 
{\displaystyle{\left| {\bf q}\right| \  \over 2(\gamma -ip)}}%
\right) \right] \quad ,
\end{eqnarray}
where 
\begin{equation}
A_{0,C_{2}}^{(^{3}S_{1})}(p)\
=-C_{2}^{(^{3}S_{1})}p^{2}\left[ 
{\displaystyle{A_{-1}^{(^{3}S_{1})}(p) \over C_{0}^{(^{3}S_{1})}}}%
\right] ^{2}\quad ,
\end{equation}
and similarly for the $^1S_0$ channel.
All other $W_{\mu \nu }^{C_{2}}=0.$  An analysis using eq.(\ref{l1beta}) shows that
the sum of $C_{2}$ and $L_{1,A}$ contributions to the hadronic tensor are $\mu$-independent.

\subsubsection{Single Potential Pion Exchange Contributions}

\begin{figure}[!t]
\centerline{{\epsfxsize=5.0in \epsfbox{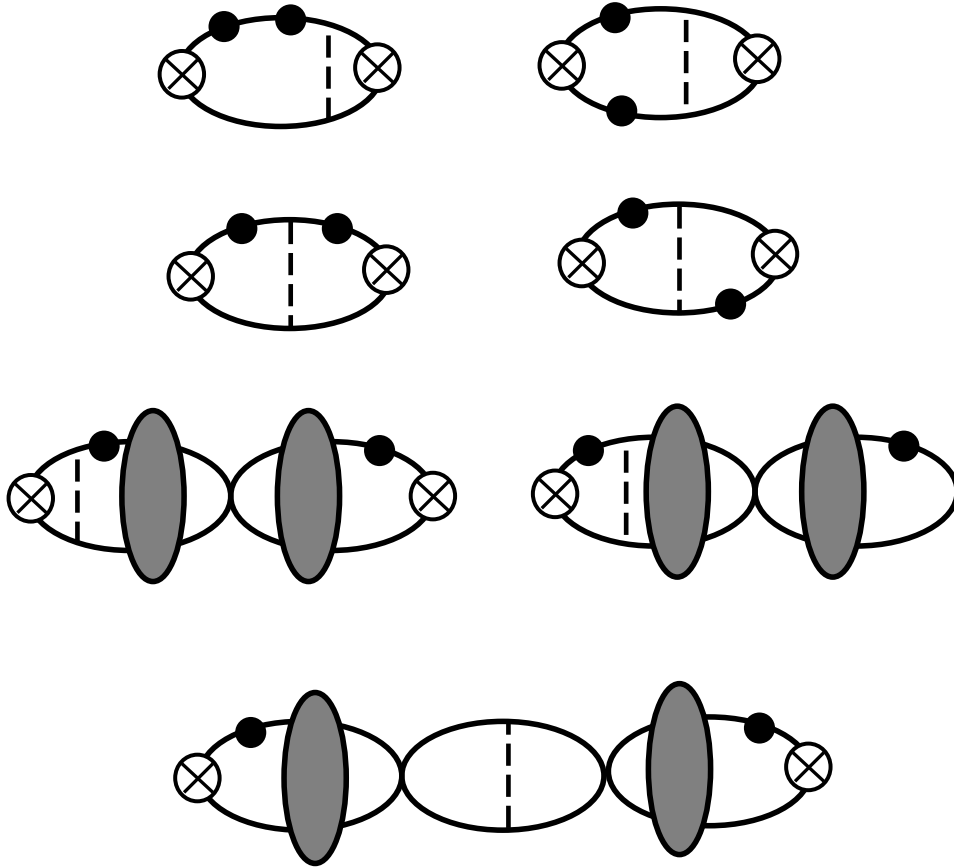}}}
\noindent

\caption{{\it Graphs from potential pion exchange that contribute to the $%
\protect\nu $-$d$ scattering hadronic tensor at NLO. The dashed lines are 
pions. The gray blobs and other features are as defined in Fig.~\ref{figLO}.
}}
\label{figPion}
\end{figure}
%

The potential-pion exchange diagrams at NLO do not yield closed-form analytical
solutions.  However, a reasonable set of
approximations could be introduced to make analytic results possible.
The diagrams contributing at NLO are shown in Fig.~\ref{figPion}.
For all diagrams 
${\bf q}$ dependence is neglected in the pion propagator. In the region $%
\omega \leq 20$ MeV that we are interested in, $\left| {\bf q}\right| \leq 40$
MeV and the error due to this approximation is estimated to be $\left\langle 
{\bf q}^{2}\right\rangle /m_{\pi }^{2}<10\%$. Further, for the diagrams in
the second row of Fig.~\ref{figPion}, we angle-averaged the pion propagator;
\begin{equation}
{\frac{1}{({\bf p}-{\bf k})^{2}+m_{\pi
}^{2}}}\rightarrow {\frac{1}{({\bf p}^{2}+{\bf k}^{2})+m_{\pi }^{2}}}\quad ,
\end{equation}
where ${\bf p}$ and ${\bf k}$ denote the nucleon momenta at the $\pi NN$
vertices. This approximation could lead to a 30\% error at this order, but
numerically the error is found to be of order 10\%. Since the terms 
neglected here are formally of order NLO, we have not performed a complete
calculation to NLO. But the error due to the approximations made 
accumulates to no more than 20\% of the NLO contribution from potential
pions, and is numerically an NNLO effect.

The structure factors can be decomposed in a manner similar to the LO result,
with
\begin{equation}
\begin{array}[b]{ll}
W_{00}^{\pi } & 
=2(C_{V}^{(0)^{2}}+C_{V}^{(1)^{2}})G_{1}+(C_{V}^{(0)^{2}}-C_{V}^{(1)^{2}})G_{2}+4C_{V}^{(0)^{2}}G_{3}\quad ,
\\ 
W_{ii}^{\pi } & =2(C_{A}^{(0)^{2}}+C_{A}^{(1)^{2}})G_{1}+%
{\displaystyle{1 \over 3}}%
(C_{A}^{(0)^{2}}-C_{A}^{(1)^{2}})G_{2}+\frac{8}{3}C_{A}^{(0)^{2}}G_{3}+\frac{%
4}{3}C_{A}^{(1)^{2}}G_{4}\quad .
\end{array}
\end{equation}
The $W_{ij(i\neq j)}^{\pi }$ components are suppressed by factors of $%
\left\langle {\bf q}^{2}\right\rangle /m_{\pi }^{2}$ compared to the
diagonal ones, so we neglect them (if they contribute at all). All other 
$W_{\mu \nu }^{\pi }=0$. The functions $G_{1,2}$ are simple modifications
of the LO functions $F_{1,2}$, given by
\begin{equation}
G_{1}=F_{1}f(p)\quad ,\qquad G_{2}=F_{2}f(p)\quad ,
\end{equation}
where 
\begin{equation}
f(p)=-%
{\displaystyle{g_{A}^{2}M_{N}m_{\pi }^{2} \over 4\pi f^{2}}}%
\left[ 
{\displaystyle{1 \over m_{\pi }+2\gamma }}%
-\ \frac{1}{p}\tan ^{-1}\left( 
{\displaystyle{p \over m_{\pi }+\gamma }}%
\right) \right] \quad .
\end{equation}
$G_{3}$ and $G_{4}$ can be further decomposed into 3 parts: 
\begin{equation}
G_{3}=G_{3}^{(a)}+G_{3}^{(b)}+G_{3}^{(c)}\quad ;\qquad
G_{4}=G_{4}^{(a)}+G_{4}^{(b)}+G_{4}^{(c)}\quad .
\end{equation}
with 
\begin{equation}
\begin{array}[b]{ll}
G_{3}^{(a)} & =F_{3}(A_{-1}^{(^{3}S_{1})}(p)\rightarrow A_{0,\pi
}^{(^{3}S_{1})}(p))\quad , \\ 
G_{4}^{(a)} & =G_{3}^{(a)}(^{3}S_{1}\rightarrow \ ^{1}S_{0})\quad ,
\end{array}
\end{equation}
where 
\begin{equation}
\begin{array}[b]{lll}
A_{0,\pi }^{(^{3}S_{1})} & =%
{\displaystyle{g_{A}^{2} \over 2f^{2}}}%
\left( 
{\displaystyle{m_{\pi }M_{N}A_{-1}^{(^{3}S_{1})}(p) \over 4\pi }}%
\right) ^{2}\bigg\{ &-\left( 
{\displaystyle{\mu +ip \over m_{\pi }}}%
\right) ^{2}\\
&&+\left[ i\tan ^{-1}\left( 
{\displaystyle{2p \over m_{\pi }}}%
\right) -%
{\displaystyle{1 \over 2}}%
\ln \left( \ 
{\displaystyle{m_{\pi }^{2}+4p^{2} \over \mu ^{2}}}%
\right) +1\right] \bigg\} \quad .
\end{array}
\end{equation}
\begin{equation}
\begin{array}[b]{lcl}
G_{3}^{(b)}&=&G_{4}^{(b)}\\
&=&  -%
{\displaystyle{g_{A}^{2}M_{N}^{2}\gamma  \over \pi ^{2}f^{2}{\bf q}^{2}(m_{\pi }^{2}+2p^{2})}}%
\tanh ^{-1}\left( 
{\displaystyle{p\left| {\bf q}\right|  \over M_{N}\nu }}%
\right) \\ 
&& \left\{ p^{2}\left[ \tan ^{-1}\left( 
{\displaystyle{\left| {\bf q}\right| -2p \over 2\gamma }}%
\right) +\tan ^{-1}\left( 
{\displaystyle{\left| {\bf q}\right| +2p \over 2\gamma }}%
\right) \right] +m_{\pi }^{2}\tan ^{-1}\left( 
{\displaystyle{\left| {\bf q}\right|  \over 2\left( \gamma +\sqrt{m_{\pi }^{2}+p^{2}}\right) }}%
\right) \right\} \quad .
\end{array}
\end{equation}
Finally,
\begin{equation}
\begin{array}{ll}
G_{3}^{(c)}= & 
{\displaystyle{4g_{A}^{2}M_{N}^{3}\gamma  \over \pi \left| {\bf q}\right| f^{2}}}%
\mathop{\rm Re}%
\bigg\{\left[ 
{\displaystyle{1 \over 8\pi ^{2}\left| {\bf q}\right| }}%
\tan ^{-1}\left( 
{\displaystyle{\left| {\bf q}\right|  \over 2\left( \gamma -ip\right) }}%
\right) \left( 
{\displaystyle{m_{\pi }^{2}\  \over m_{\pi }+2\gamma }}%
+\mu +ip\right) +h(\gamma ,p)+h(-ip,i\gamma )\right] \\ 
& \tan ^{-1}\left( 
{\displaystyle{\left| {\bf q}\right|  \over 2\left( \gamma -ip\right) }}%
\right) iA_{-1}^{(^{3}S_{1})}(p)\bigg\}\quad , \\ 
G_{4}^{(c)}= & G_{3}^{(c)}(^{3}S_{1}\rightarrow \ ^{1}S_{0})\quad ,
\end{array}
\end{equation}
where 
\begin{eqnarray}
h(\gamma,p) &=&%
{\displaystyle{-m_{\pi }^{2} \over 16\pi ^{2}(m_{\pi }+\gamma )}}%
\left\{ -%
{\displaystyle{1 \over \gamma -ip}}%
+%
{\displaystyle{m_{\pi }+\gamma  \over \gamma ^{2}+p^{2}}}%
\ln \left[ 
{\displaystyle{m_{\pi }+2\gamma  \over m_{\pi }+\gamma -ip\ }}%
\ \right] \right.  \nonumber \\
&&\left. +\frac{i}{\left| {\bf q}\right| }\ln \left[ 
{\displaystyle{{\bf q}^{2}+4(\gamma ^{2}+p^{2}+\left| {\bf q}\right| p) \over -{\bf q}^{2}+4(\gamma ^{2}+p^{2}+i\left| {\bf q}\right| \gamma )}}%
\ \right] \right\} \quad .
\end{eqnarray}

\subsubsection{$W_{\protect\mu \protect\nu }^{M_{Z}}$ Contributions}

\begin{figure}[!t]
\centerline{{\epsfxsize=5.0in \epsfbox{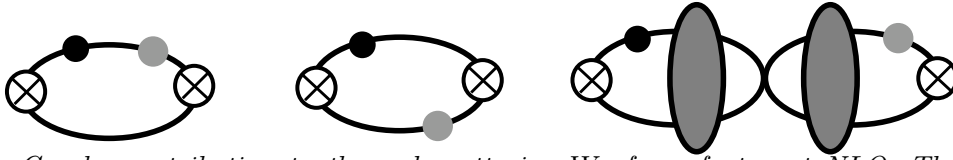}}}
\noindent

\caption{{\it Graphs contributing to the $\protect\nu $-$d$ scattering 
$W_3$ form factor at NLO. The dark solid circles correspond to insertions  
of the leading weak current operators, while the gray solid
circles correspond to insertions of the next-to-leading weak magnetic
moment operators. The gray blobs and other features are as defined
in Fig.~\ref{figLO}.}}
\label{figW3}
\end{figure}
%

The diagrams with one LO single-nucleon weak current and one NLO weak magnetic moment coupling, as
shown in Fig.~\ref{figW3}, contribute to the antisymmetric part of the NLO
hadronic tensor 
\begin{equation}
W_{ij}^{M_{Z}}=-iW_{3}^{M_{Z}}\varepsilon _{0ijk}%
{\displaystyle{q^{k} \over M_{d}}}%
\quad ,
\end{equation}
with all other $W_{\mu \nu }^{M_{Z}}=0$. It is straightforward to
show that
\begin{eqnarray}
W_{3}^{M_{Z}} &=&-4(C_{A}^{(0)}C_{M}^{(0)}+C_{A}^{(1)}C_{M}^{(1)})F_1-%
{\displaystyle{2 \over 3}}%
(C_{A}^{(0)}C_{M}^{(0)}-C_{A}^{(1)}C_{M}^{(1)})F_2  \nonumber \\
&&-\frac{16}{3}C_{A}^{(0)}C_{M}^{(0)}F_3-\frac{8}{3}%
C_{A}^{(1)}C_{M}^{(1)}F_4\quad ,
\end{eqnarray}
where
\begin{equation}
C_{M}^{(0)}=-2\sin ^{2}\vartheta _{W}\kappa ^{(0)}-\frac{1}{2}\mu _{S}\quad
,\quad C_{M}^{(1)}=(1-2\sin ^{2}\vartheta _{W})\kappa ^{(1)}\quad .
\end{equation}
Recall that at LO, $W_3$ vanished.  As such, this NLO contribution
is the leading contribution to the difference between neutrino and antineutrino
scattering from the deuteron.

\section{Antineutrino-Deuteron Charged Current Inelastic Scattering}

The differential cross-section for the process
\begin{equation}
\overline\nu+d\rightarrow e^+ +n+n
\end{equation}
takes the same form as eq.~(\ref{dsig}).  There are, in principle, corrections 
due to the finite positron mass $m_e$, but their impact on
the total cross-section is negligible,
even at threshold (they could, however, affect the angular distribution).  We must 
also modify the phase space bounds
to include the effects of the positron's mass, as well as the neutron--proton
mass splitting $\delta m=m_{n}-m_{p}$. The modified bounds are 
\begin{gather}
\text{Max}\left[ -1,-{\displaystyle{\frac{4M_{N}(\nu -B-\delta m)-\omega
^{2}-\omega ^{\prime 2}+m_{e}^{2}}{2\omega \sqrt{\omega ^{\prime 2}-m_{e}^{2}%
}\ }}}\right] \leq \cos \theta \leq 1\quad ,  \nonumber \\
m_{e}\leq \omega ^{\prime }\leq \omega -2(M_{N}-\sqrt{M_{N}^{2}-M_{N}(B+%
\delta m)})\ .
\end{gather}

For the most part, however, the primary difference between the neutral
current and charged current cases is the fact that the charged current
processes are purely isovector. As a result, the charged current structure factors
can be obtained from the neutral current structure factors with the
simple substitutions:

\begin{eqnarray}
&& C_{V}^{(0)} = 0\quad , \quad C_{V}^{(1)}={{%
{\displaystyle{\left| V_{ud}\right|  \over \sqrt{2}}}%
}}\quad ,\quad  \nonumber \\
&& C_{A}^{(0)}=0\quad , \quad C_{A}^{(1)}={{%
{\displaystyle{\left| V_{ud}\right|  \over \sqrt{2}}}%
}}g_{A}\quad ,  \nonumber \\
&& C_{M}^{(0)}=0\quad ,\quad C_{M}^{(1)}=\sqrt{2}\left| V_{ud}\right| \kappa
^{(1)}\quad ,  \nonumber \\
&& L_{2,A}=0\quad ,\quad 
L_{1,A}\rightarrow \sqrt{2}L_{1,A}\left| V_{ud}\right| \quad , \nonumber \\
&&\nu \rightarrow \nu -\delta m\quad ,
\end{eqnarray}
where we use $\left| V_{ud}\right| =0.975$ for this CKM matrix element.

\section{Numerical Results}

We have presented analytic expressions of the differential cross-sections $%
{\displaystyle{d^{2}\sigma  \over d\omega ^{^{\prime }}d\Omega }}%
$ for $\nu d\rightarrow \nu np$, $\bar{\nu}d\rightarrow \bar{\nu}np$ and $%
\bar{\nu}d\rightarrow e^{+}nn$ processes in the previous sections. These
expressions have the correct elastic scattering limits and their threshold
behaviors are consistent with \cite{SSWst}. We can now consider the numerical
results for the total cross-sections of these processes.  In calculating
cross-sections for the different channels, we will include the isospin 
splitting (charge
dependence) of the strong interaction in the $^1S_0$ channel, but it can
be neglected in the weak interaction currents as a higher-order effect.

In the NLO calculations, four parameters ($\Delta s$, $\mu _{s}$, $L_{1,A}$,
and $L_{2,A})$, which are not well constrained, contribute with different
significance. To demonstrate their numerical effect we express the NC 
cross-sections at $\omega =10$ MeV as 
\begin{equation}
\begin{array}{ll}
\sigma (\nu (\bar{\nu})d\rightarrow \nu (\bar{\nu})np) & =0.999 \pm 0.026+
0.013 L_{1,A}\\
& +10^{-5} \Delta s (\pm 0.5 \pm 1.2 \mu_s +6.3 \Delta s-4.6 L_{2,A}) \quad ,
\end{array}
\end{equation}
where $L_{1,A}$ and $L_{2,A}$ are in units of fm$^{3}.$ It is clear, immediately,
that $\Delta s$ and $\mu _s$ contribute less than 1\% to the total
cross-section. Furthermore, since
dimensional analysis suggests $L_{1,A}$ and $L_{2,A}$ are of order 
\begin{equation}
\left| L_{1,A}\right| \approx \left| L_{2,A}\right| \approx {%
{\displaystyle{4\pi  \over M}}%
{\displaystyle{1 \over \mu ^{2}}}%
}\sim 5\ {\rm fm}^{3}\quad 
\label{dimanal}
\end{equation}
at $\mu=m_\pi$, we see that $L_{1,A}$ could contribute at the $10\%$ level but 
that $L_{2,A}$ would 
contribute far less than 1\% to the total cross-section. 
This agrees with the $^{3}S_{1}\rightarrow
\,^{3}S_{1}$ suppression mentioned in section \ref{thresh}. In this
computation, $L_{1,A}$ is kept as a free parameter,  $\Delta s=-0.17,$
$\mu _{s}=0$, and $L_{2,A}=0.$ The error from this choice is estimated to
be less than 1\%.

\begin{figure}[!t]
\centerline{{\epsfxsize=4. in \epsfbox{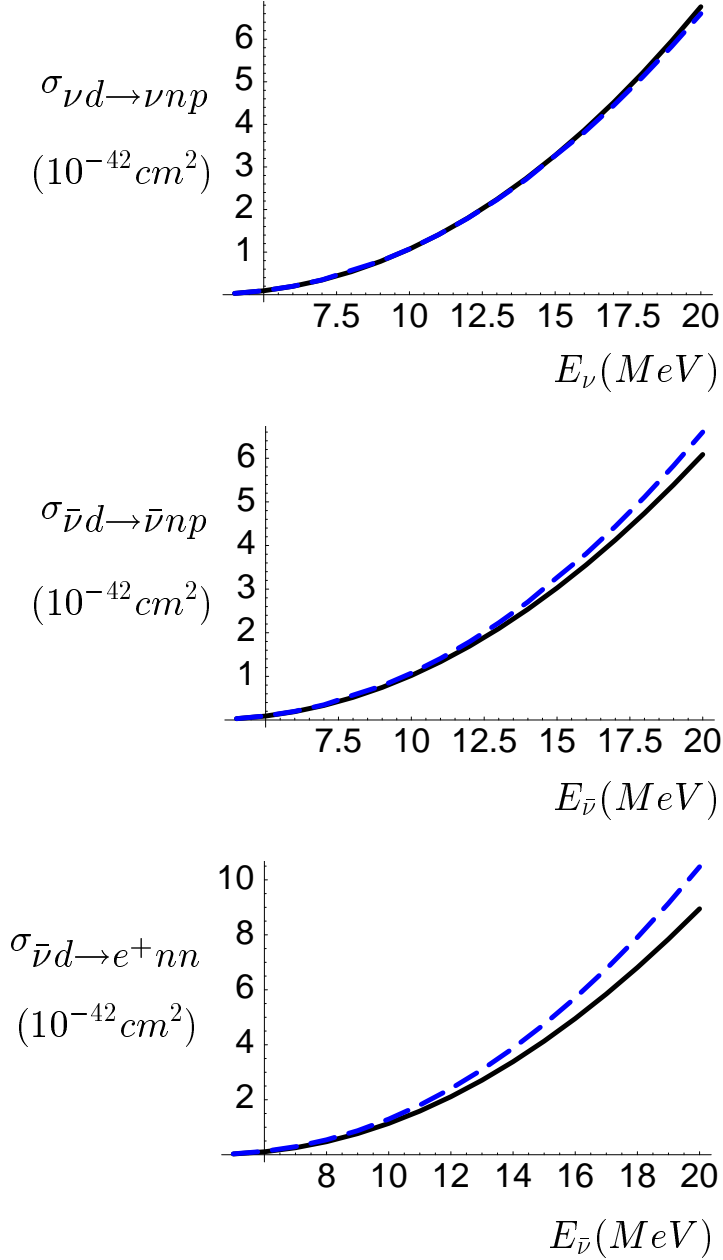}}}
\noindent

\caption{{\it Cross-sections for
$\protect\nu d\rightarrow \protect\nu np$ (top graph), $\bar{\protect%
\nu}d\rightarrow \bar{\protect\nu}np$ (middle graph) and $%
\bar{\protect\nu}d\rightarrow e^{+}nn$ (bottom graph) shown as functions of incident $\protect\nu (\bar{\protect\nu})$ energy. The
solid curves correspond to the average of YHH and KN results while the
dashed curves correspond to the LO results in EFT. Note that
at LO, there is no difference in EFT results for $\protect\nu d\rightarrow \protect\nu np$
and $\bar{\protect\nu}d\rightarrow \bar{\protect\nu}np$ cross-sections.
 }}
\label{figBCLO}
\end{figure}
%

The LO results are parameter free, and are shown in Fig.~\ref{figBCLO}
together with potential model results. 
With respect to the
two most modern potential model calculations used, Kubodera {\it et\ al.}\/'s
results (KN\cite{KN})\footnote{We note that while these calculations
are presented in Kubodera and Nozawa of ref.~\cite{KN}, they are actually the unpublished work
of Kohyama and Kubodera, also listed in ref.~\cite{KN}.} 
are persistently larger than Ying {\it et al.}\/'s    
results (YHH\cite{YHH}) by 5-10\% in all channels.  Thus, we average the KN and YHH
calculations in each channel, before comparing them to our LO results. 
First, consider the NC channels. At LO there is no difference in our calculation between
$\nu d\rightarrow\nu np$ and $\overline\nu d\rightarrow \overline\nu np$
cross-sections. We see then that our LO result
lies within 10\% of the (averaged) potential model results.
A similar level of
agreement is seen in the comparison of our LO calculation of 
$\overline\nu d \rightarrow e^+ nn$ to (the averaged) potential model calculations.

\begin{table}[!tbp]
\begin{tabular}[h]{ccccccc}
& \multicolumn{2}{c}{$\nu d\to\nu np$} & \multicolumn{2}{c}{$\bar{\nu} d\to 
\bar{\nu} np$} & \multicolumn{2}{c}{$\bar{\nu} d\to e^+ nn$} \\ 
$E_{\nu(\bar{\nu})}$ MeV & a & b & a & b & a & b \\ \hline
4 & 0.0288 & 0.000299 & 0.0283 & 0.000299 &  &  \\ 
5 & 0.0888 & 0.000965 & 0.0869 & 0.000965 & 0.0264  & 0.000268 \\ 
6 & 0.188 & 0.00212 & 0.183 & 0.00212 & 0.111 & 0.00120 \\ 
7 & 0.330 & 0.00381 & 0.319 & 0.00381 & 0.262 & 0.00299 \\ 
8 & 0.516 & 0.00609 & 0.496 & 0.00609 & 0.484 & 0.00576 \\ 
9 & 0.747 & 0.00899 & 0.714 & 0.00899 & 0.780 & 0.00960 \\ 
10 & 1.02 & 0.0125 & 0.973 & 0.0125 & 1.15 & 0.0146 \\ 
11 & 1.35 & 0.0167 & 1.27 & 0.0167 & 1.59 & 0.0207 \\ 
12 & 1.72 & 0.0216 & 1.61 & 0.0216 & 2.11 & 0.0281 \\ 
13 & 2.14 & 0.0271 & 2.00 & 0.0271 & 2.70 & 0.0368 \\ 
14 & 2.61 & 0.0334 & 2.42 & 0.0334 & 3.35 & 0.0467 \\ 
15 & 3.12 & 0.0404 & 2.88 & 0.0404 & 4.08 & 0.0580 \\ 
16 & 3.69 & 0.0480 & 3.38 & 0.0480 & 4.88 & 0.0706 \\ 
17 & 4.30 & 0.0564 & 3.92 & 0.0564 & 5.74 & 0.0846 \\ 
18 & 4.97 & 0.0656 & 4.50 & 0.0656 & 6.68 & 0.0999 \\ 
19 & 5.68 & 0.0754 & 5.12 & 0.0754 & 7.68 & 0.117 \\ 
20 & 6.45 & 0.0860 & 5.77 & 0.0860 & 8.74 & 0.135
\end{tabular}
\caption{NLO $\protect\nu d$ and $\bar{\protect\nu}d$ inelastic scattering
results in EFT with the cross-sections written in the form $\protect\sigma 
=(a+bL_{1,A})\times 10^{-42}{\rm cm}^{2}$ with $L_{1,A}$ in units of ${\rm fm}^{3}$.
}
\label{table1}
\end{table}

\begin{figure}[!t]
\centerline{{\epsfxsize=5 in \epsfbox{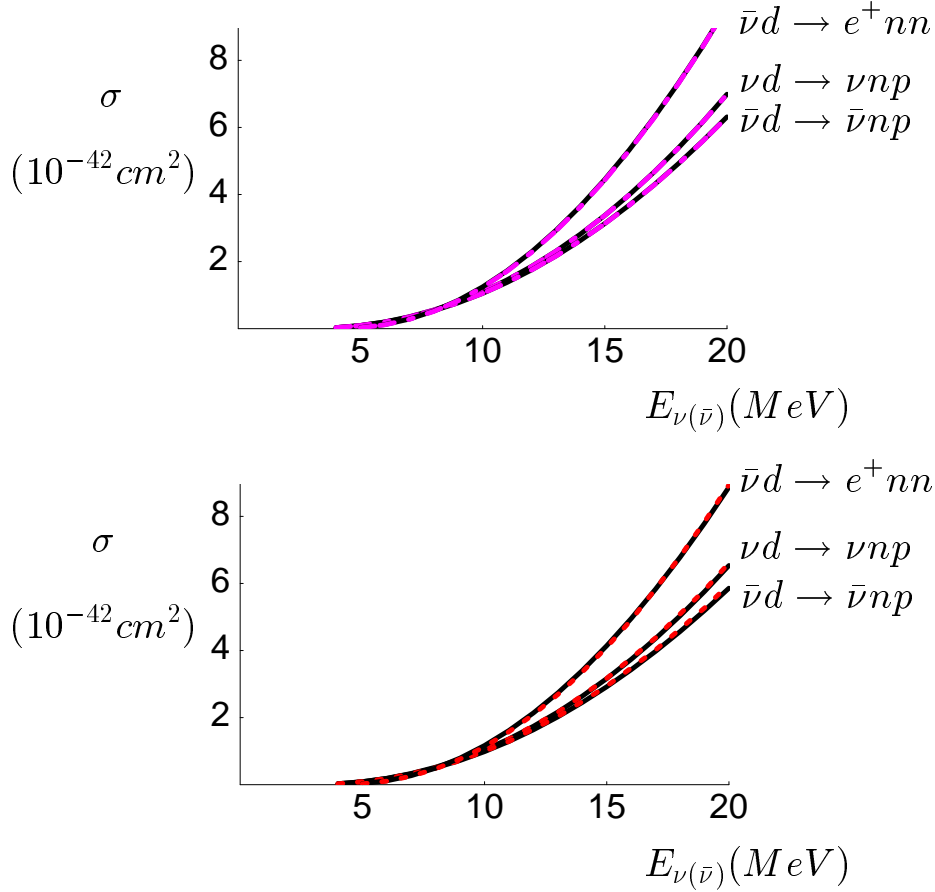}}}
\noindent

\caption{{\it Inelastic $\protect\nu (\bar{\protect\nu})d$ cross-sections
shown as functions of incident $\protect\nu (\bar{\protect\nu})$ energy. The
solid curves in the upper graph are KN results while the dot-dashed curves,
which lie right on top of the solid curves, are NLO results in EFT with $%
L_{1,A}=6.3\ {\rm fm}^{3}$. The solid curves in the lower graph are YHH results
while the dashed curves, which also lie right on top of the solid curves,
are NLO results in EFT with $L_{1,A}=1.0\ {\rm fm}^{3}$. In both graphs, the highest
curves are $\bar{\protect\nu}d\rightarrow e^{+}nn$ cross-sections while the
middle and lower curves are $\protect\nu d\rightarrow \protect\nu np$ and $%
\bar{\protect\nu}d\rightarrow \bar{\protect\nu}np$ cross-sections
respectively. }}
\label{figBCNLO}
\end{figure}

Our cross-sections at NLO will have one free parameter, $L_{1,A}$. 
To facilitate
study of the effects of this two-body contribution, we tabulate 
the NLO cross-sections as
functions of $L_{1,A}$ in Table \ref{table1}. It is not surprising that
we can reproduce the total cross-section results of KN and YHH in all 
three channels to a high degree of 
accuracy by choosing $L_{1,A}$ to be 6.3 fm$^{3}$ and 1.0 fm$^{3}$
respectively (see Fig.~\ref{figBCNLO}). This would imply that the 5-10\%
systematic errors in these two potential model calculations are due solely
to different assumptions made about the short distance physics. It also
suggests that EFT is a perfect tool to study the $\nu d$ breakup processes
because, once $L_{1,A}$ is fixed by one experiment, predictions can be made
in other channels such as the solar fusion process $pp\rightarrow
de^{+}\nu$, and supernova short term cooling processes such as $%
np\rightarrow np\nu \overline{\nu }.$ One experiment which could, in
principle, constrain $L_{1,A}$ is parity-violating $\overrightarrow{e}%
d\rightarrow enp$. The SAMPLE II collaboration~\cite{SAMPLEII} is measuring this reaction at
MIT-Bates with the intent of studying the strange magnetic moment, $\mu _{s}$.
Their kinematics are tuned for that purpose, and are not well-suited for an
extraction of $L_{1,A}$. It would be useful to explore future possibilities
that optimize an extraction of the isovector axial coupling, $L_{1,A}$.


%

\begin{figure}[!t]
\centerline{{\epsfxsize=4.0 in \epsfbox{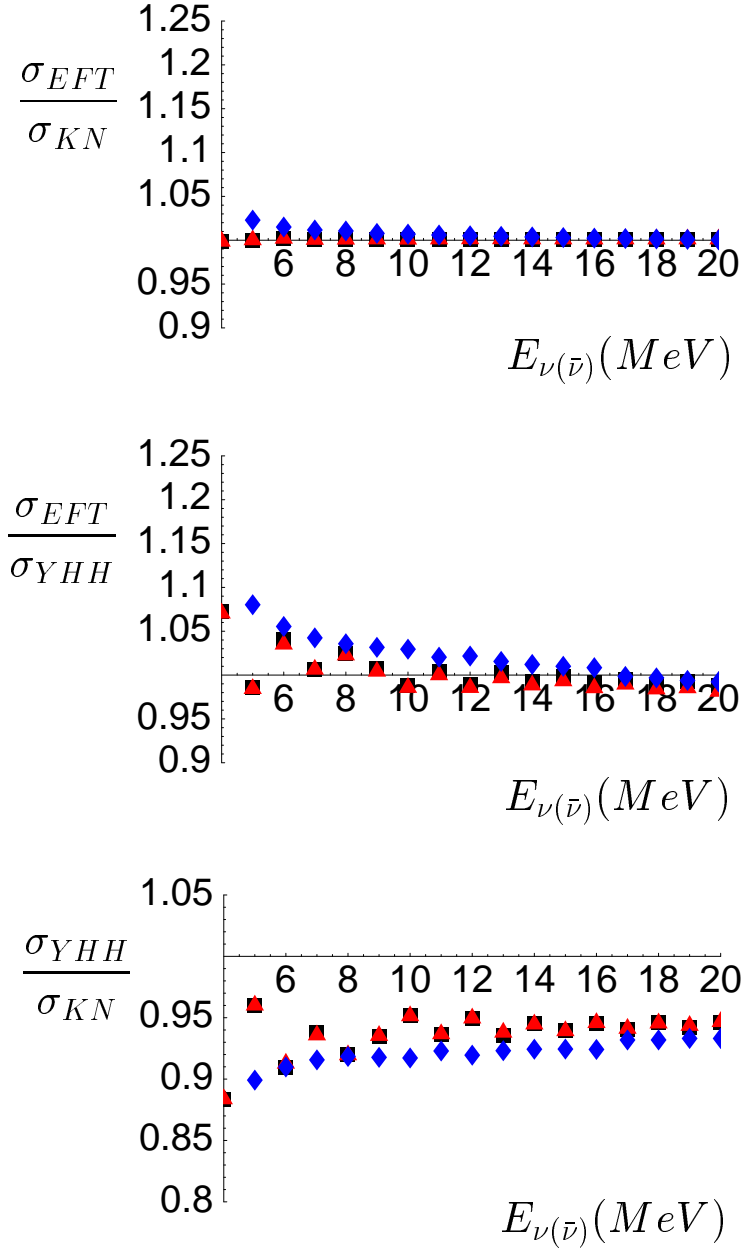}}}

\noindent

\caption{{\it Ratios between different calculations of the inelastic $%
\protect\nu(\bar{\protect\nu})d$ cross-sections shown as functions of
incident $\protect\nu (\bar{\protect\nu})$ energy. The first graph
corresponds to $\protect\sigma_{EFT}/\protect\sigma_{KN}$ while the second
and third graphs correspond to $\protect\sigma_{EFT}/\protect\sigma_{YHH}$
and $\protect\sigma_{YHH}/\protect\sigma_{KN}$ respectively. In all graphs,
boxes are the $\protect\nu d\to\protect\nu np$ ratios, while triangles and
diamonds are the $\bar{\protect\nu} d\to\bar{\protect\nu} np$ and $\bar{%
\protect\nu} d\to e^+ nn$ ratios. }}
\label{figBCTKKYHHratio}
\end{figure}

To make a more detailed comparison with potential model calculations, we
choose the same $L_{1,A}$ values as mentioned above and plot the ratios
between different calculations of the three $\nu(\overline\nu)d$ reaction
channels in Fig.~\ref{figBCTKKYHHratio}. We find that the agreement is
excellent between our calculations and the potential model results of KN (within
2\%),
and that we agree with the results of YHH to within 10\% (the actual agreement
is likely better than this -- the fluctuations in YHH's calculation may exaggerate
the differences somewhat).
In making these comparisons, we should recall that we have taken into account
isospin splitting in the $^1S_0$ rescattering amplitudes for $nn$ and $np$ final states.
The details can be found in the Appendix.
Further, for $np$ final-states, our fit for the $^1S_0$ channel is not as good as for the $^3S_1$ channel. 
However, it begins to deviate from the measured phase shifts only above
relative momenta of 50~MeV.  The sensitivity of the calculation to these
momenta is suppressed by the factor of $\omega^{\prime}|{\bf k}^\prime|$ in the 
differential cross-section for NC and CC scattering (eq.~(\ref{dsig})).  
$\omega^{\prime}\propto(p_{max}^2-p^2)$ means that
the calculation is most sensitive to lower relative momenta.  

%

\begin{figure}[!t]
\centerline{{\epsfxsize=4.0 in \epsfbox{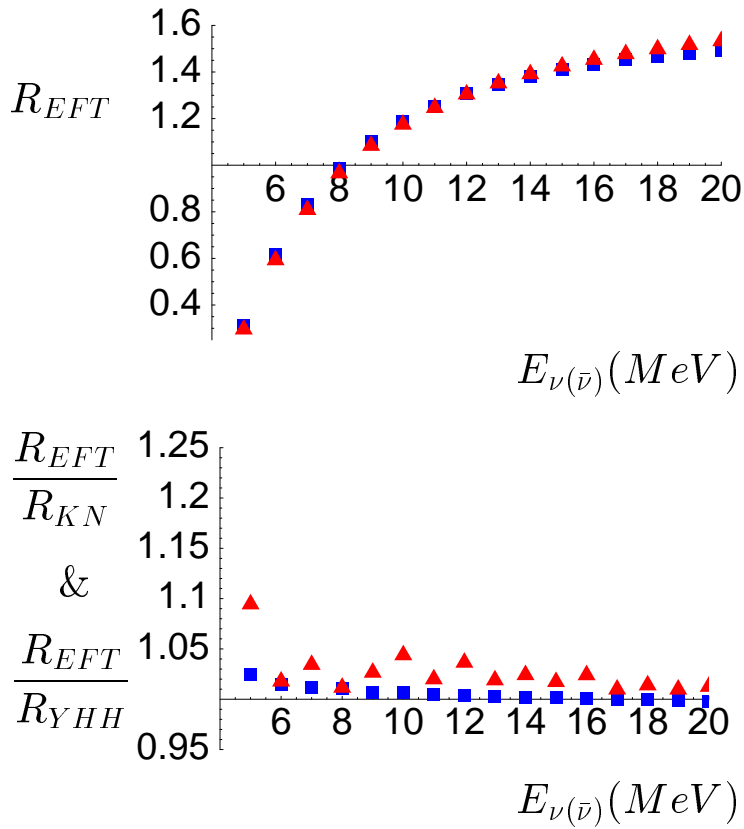}}}
\noindent
\caption{{\it $\bar{\protect\nu} d$ scattering CC to NC cross-section ratios
$R$ shown as functions of incident $\bar{\protect\nu}$ energy. The upper
graph corresponds to the NLO EFT result with $L_{1,A}=-20~{\rm fm}^3$
(boxes) and $40~{\rm fm}^3$ (triangles). The lower graph correspond to the
ratios $R_{EFT}/R_{KN}$ (boxes) and $R_{EFT}/R_{YHH}$ (triangles) with
input $L_{1,A}=3.7\ {\rm fm}^3$ for the evaluation of $R_{EFT}$.
}}
\label{figCCNCratio}
\end{figure}
%

Another interesting quantity is the ratio $R$ between CC and 
NC scattering cross-sections,
\begin{equation}
R \equiv {\sigma_{CC} \over \sigma_{NC}}\quad .
\end{equation}
This ratio is zero at threshold ($E_{\overline\nu}=4.03$~MeV). Above
threshold, it is expected to be insensitive to the treatment or
modeling of
short distance physics and, in fact, 
changes by only 
$5\%$ at NLO above 5~MeV when $L_{1,A}$ varies over a wide range, from $-20$ to 40 fm$^{3}$.
Choosing $L_{1,A}=3.7$~fm$^3$, our values of $R$ agree well with those 
obtained by KN and YHH over the full range of neutrino energies studied
(Fig.~\ref{figCCNCratio}). 

One of the strengths of effective field theory is that it provides us with the
ability to quantify theoretical uncertainty.  It remains for us to
quantify the precision of the EFT calculations presented
here. Naive power counting tells us that the NLO calculation has a $10\%$ uncertainty
from higher order corrections, even if we were able to fit the
counterterm $L_{1,A}$ from experimental data (such as the
$\overrightarrow{e} d$
scattering process mentioned earlier). This 
estimate is consistent with the observation that setting $L_{1,A}$ to
$+6$, $0$, and $-6$ fm$^3$ (sizes suggested from the dimensional analysis of 
eq.~(\ref{dimanal})) corresponds to the NLO contribution being $10$, $20$ and 
$30\%$ (respectively) of the LO contribution. As the NLO contribution is naively
a $30\%$ effect, the NNLO effect could be estimated to be $\sim 10\%$.
However, if the sign of $L_{1,A}$ is positive, as preferred by our comparison
to the potential model calculations of KN and YHH, the 
uncertainty from NNLO could be only $\sim 4\%$. This latter scenario
would also explain
why we can fit the potential model results at (in general) the few percent 
level, given that higher order contributions are partially incorporated in 
very different fashions in different calculations. The size of NNLO effects
will be studied further~\cite{BCnopi} using EFT without pions~\cite{CRS}.    

\section{Conclusions}

We have presented analytic differential cross-sections for elastic and
inelastic neutrino deuteron scattering processes using effective field
theory. For elastic scattering, the deuteron axial form factor arising from
strange matrix elements, and the deuteron strange magnetic moment form factor are
computed to NLO with two-body current dependence. For inelastic scattering,
two neutral current processes $\nu d\rightarrow \nu np$, $\overline{\nu }%
d\rightarrow \overline{\nu }np$ and one charge current process $\overline{%
\nu }d\rightarrow e^{+}nn$ are computed to NLO with an isovector axial
two-body matrix element whose value is yet to be fixed by experiment.
Potential model calculations done by Kubodera {\it et al.}~and Ying
{\it et al.}~are reproduced, with a high degree of accuracy, by choosing 
different values for the
two-body matrix element.  This implies that the differences between the
two potential model calculations lie in their treatment of short distance
physics. The charged current to
neutral current $\overline{\nu }d$ cross-section ratio is confirmed to be
insensitive to short distance physics, and the same ratio is obtained by
potential models and this calculation within our intrinsic uncertainties 
(conservatively estimated to be 5\%) for the full range of incident neutrino
energies studied, up to 20~MeV. 
The two-body matrix element could be fixed using the 
parity-violating process $\overrightarrow{e}d\rightarrow e np$.

There still remains the need to calculate the other charged current
process, $\nu d\rightarrow e^- pp$, in EFT.  This is the primary reaction
channel at SNO and issues of its precision and accuracy should not be
simply inferred from the processes studied here.  The complications 
introduced because of coulomb interactions in the final state make it
more appropriate to discuss this process elsewhere~\cite{BCK}, along with
details of the angular distributions of the charged-current reactions~\cite{VogelBeacom}.
\vskip2in \centerline{\bf ACKNOWLEDGMENTS} \bigskip We would like to thank
Martin Savage, Doug Beck, Elizabeth Beise, Wick Haxton, Sanjay Reddy, Hamish Robertson, 
 and Roxanne Springer for useful discussions. This work is supported, in part, by
the U.S. Dept. of Energy under Grant No.\ DE-FG03-97ER4014.
M.B.\ is supported by a grant from the Natural Sciences and Engineering
Research Council of Canada.

\vskip0.5cm \centerline{\bf APPENDIX} \bigskip

This appendix summarizes how the $NN$ scattering strong interaction
parameters can be fit to phase shift data by matching onto the effective range
expansion (ERE). 
The $NN$ scattering amplitude ${\cal A}$ has a power series expansion in $Q$; $%
{\cal A}={\cal A}_{-1}+{\cal A}_{0}+...$, where the subscripts denote the
powers in $Q$. ${\cal A}_{-1}$ and ${\cal A}_{0}$ have been calculated in 
\cite{KSW}. 
The standard parameterization of the $NN$ scattering amplitude is given by 
\begin{equation}
{\cal A}=\frac{4\pi }{M_{N}}\frac{1}{p\cot \delta -ip}\quad .
\label{ampdelta}
\end{equation}
where $p$ is the relative momentum and $\delta$ is the phase shift.
The effective range expansion is a power series expansion of
$p\cot\delta$, yielding
\begin{equation}
p\cot \delta =-\frac{1}{a}+\frac{1}{2}%
r_{0}p^{2}+\cdots \quad ,
\end{equation}
where $a$ is the scattering length and $r_{0}$ is the effective range. 
For the $^{1}S_{0}$ channel, we can rewrite eq.~(\ref{ampdelta}) as 
\begin{eqnarray}
{\cal A}^{ERE} &=&-\frac{4\pi }{M_{N}}\frac{1}{\left( \frac{1}{%
a^{(^{1}S_{0})}}+ip\right) }\left[ 1+\frac{1}{2}\frac{%
r_{0}^{(^{1}S_{0})}p^{2}}{\left( \frac{1}{a^{(^{1}S_{0})}}+ip\right) }%
+\cdots \right] \quad   \nonumber \\
&=&{\cal A}_{-1}^{ERE}+{\cal A}_{0}^{ERE}+...\quad ,
\end{eqnarray}
where we have performed a $Q$-expansion.
At LO, we can relate the coefficient $C_0^{^1S_0}$
to effective range parameters through equating
\begin{equation}
{\cal A}_{-1}^{{}}(p)={\cal A}_{-1}^{ERE}(p)
\end{equation}
yielding
\begin{equation}
C_{0}^{(^{1}S_{0})}(\mu )=-\frac{4\pi }{M_{N}}\frac{1}{\left( \mu -{1 \over 
a^{^1S_0}} \right) }\quad .
\end{equation}
At NLO, ${\cal A}_{0}$ does not have the same $p$ dependence as ${\cal A}%
_{0}^{ERE}.$ The NLO matching is performed near $p=0$. 
As $p\rightarrow 0,$%
\begin{equation}
{\cal A}^{ERE}_{0}(p)\rightarrow -\frac{2\pi }{M_{N}}%
a^{(^{1}S_{0})^{2}}r_{0}^{(^{1}S_{0})}p^{2}\quad .
\end{equation}
The matching yields
\begin{equation}
D_{2}^{(^{1}S_{0})}(m_{\pi })=0\ \quad ,
\end{equation}
and
\begin{equation}
C_{2}^{(^{1}S_{0})}(m_{\pi })=\frac{2\pi }{M_{N}}\frac{r_{0}^{(^{1}S_{0})}}{%
\left( m_{\pi }-\frac{1}{a^{(^{1}S_{0})}}\right) ^{2}}-\frac{g_{A}^{2}\left(
3a^{(^{1}S_{0})^{2}}m_{\pi }^{2}-8a^{(^{1}S_{0})}m_{\pi }+6\right) }{%
6f^{2}a^{(^{1}S_{0})^{2}}m_{\pi }^{2}\left( m_{\pi }-\frac{1}{a^{(^{1}S_{0})}%
}\right) ^{2}}\ \quad . 
\end{equation}
The ERE parameters 
\begin{equation}
\begin{array}[!t]{ll}
a_{np} = -23.714 \pm 0.013\text{ fm}\quad&,\quad
r_{0_{np}} = 2.73 \pm 0.03 \text{ fm}\quad , \nonumber \\
a_{nn} = -18.5 \pm 0.4\text{ fm}\quad&,\quad
r_{0_{nn}} = 2.80 \pm 0.11 \text{ fm}\quad, 
\end{array}
\end{equation}
are different between n-p and n-n systems \cite{Coon}.
Taking into account this observed violation of charge independence, we find
for the n-p system   
\begin{equation}
C_{0}^{(^{1}S_{0})}(m_{\pi })=-3.56\text{ fm}^{2}\quad ,\quad
D_{2}^{(^{1}S_{0})}(m_{\pi })=0\text{ fm}^{4}\quad ,\quad
C_{2}^{(^{1}S_{0})}(m_{\pi })=2.79\text{ fm}^{4}\quad ,  \label{nsol1s0}
\end{equation}
while for the n-n system 
\begin{equation}
C_{0}^{(^{1}S_{0})}(m_{\pi })=-3.50\text{ fm}^{2}\quad ,\quad
D_{2}^{(^{1}S_{0})}(m_{\pi })=0\text{ fm}^{4}\quad ,\quad
C_{2}^{(^{1}S_{0})}(m_{\pi })=2.71\text{ fm}^{4}\quad .
\end{equation}

\begin{figure}[!t] 
\centerline{{\epsfxsize=4.0 in \epsfbox{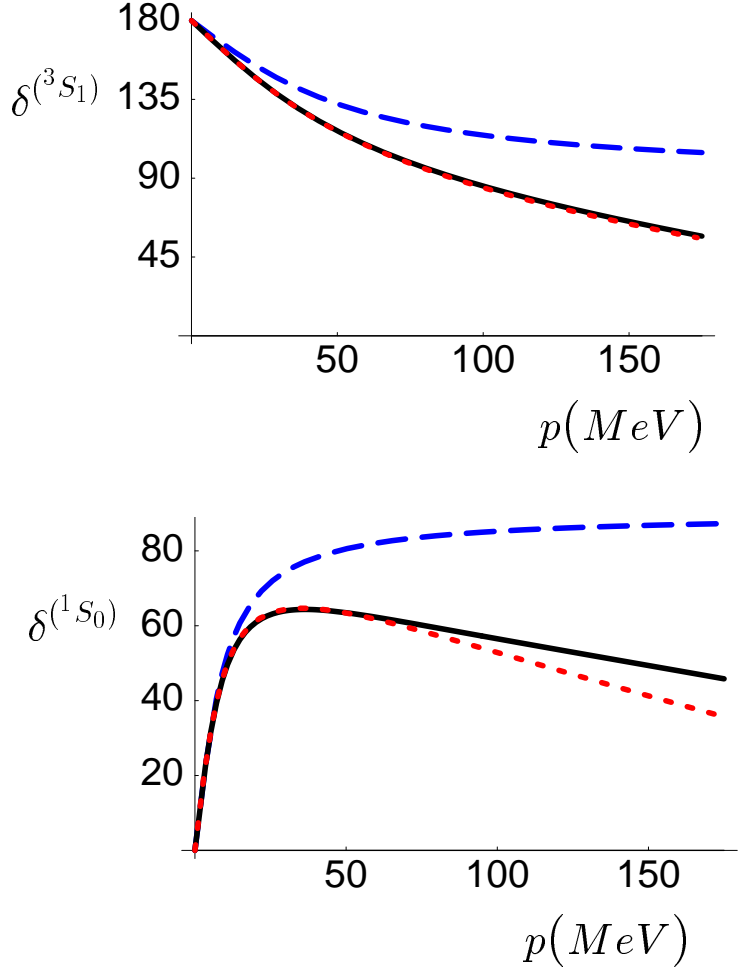}}}  
\noindent
\caption{{\it n-p scattering phase shifts (measured in degrees) for the $ ^3S_1$ (upper graph) and $ ^1S_0$ 
channels (lower graph). The dashed curves are the LO results and the solid curves are 
the NLO results from eqs.(\ref{nsol3s1},\ref{nsol1s0}). The dotted curves are 
the results of the Nijmegen partial wave analysis. 
}}
\label{Fig:phaseshifts}
\end{figure}

For the $^{3}S_{1}$ system, $p\cot\delta$ is usually
expanded around the deuteron pole, 
\begin{equation}
p\cot \delta =-\gamma +\frac{1}{2}r_{0}^{(^{3}S_{1})}(p^{2}+\gamma
^{2})+\cdots \quad ,  \label{ERE}
\end{equation}
such that we can rewrite eq.(\ref{ampdelta}) as 
\begin{eqnarray}
{\cal A}^{ERE}
&=&\frac{4\pi }{M_{N}}\frac{1}{-\gamma -ip\ }\left[ 1+\frac{1}{2}%
r_{0}^{(^{3}S_{1})}(\gamma -ip)+\cdots \right]  \nonumber \\
&=&{\cal A}_{-1}^{ERE}+{\cal A}_{0}^{ERE}+...\quad ,
\end{eqnarray}
after performing a $Q$-expansion. 
The LO matching yields
\begin{equation}
C_{0}^{(^{3}S_{1})}(\mu )=-\frac{4\pi }{M_{N}}\frac{1}{\left( \mu -\gamma
\right) }\quad .
\end{equation}
At NLO, the matching is performed at $p=0$ and $i\gamma$, so that
the scattering length is reproduced and the residue of the
deuteron pole is not changed in the NLO EFT amplitude. That yields two 
conditions 
\begin{eqnarray}
{\cal A}_{-1}^{{}}(0)+{\cal A}_{0}^{{}}(0)=-\frac{4\pi a^{(^{3}S_{1})}}{M_{N}%
}\quad ,  \nonumber \\
{\cal A}_{0}^{{}}(i\gamma ) ={\cal A}_{0}^{ERE}(i\gamma )=0\quad .
\end{eqnarray}
Matching at $\mu =m_{\pi }$, these two conditions can be solved for the coefficients
$D_2^{^3S_1}$ and $C_2^{^3S_1}$
\begin{equation}
D_{2}^{(^{3}S_{1})}(m_{\pi })=-\frac{4\pi }{M_{N}}\frac{\gamma \left(
1-a^{(^{3}S_{1})}\gamma \right) }{m_{\pi }^{2}\left( m_{\pi }-\gamma \right)
^{2}}\quad ,  \label{sol3s12}
\end{equation}
and 
\begin{equation}
C_{2}^{(^{3}S_{1})}(m_{\pi })=\frac{1}{2\gamma ^{2}\ }\left[ \left(
2D_{2}^{(^{3}S_{1})}(m_{\pi })m_{\pi }^{2}+\frac{g_{A}^{2}}{f^{2}}\right) +%
\frac{g_{A}^{2}m_{\pi }^{2}}{f^{2}\left( m_{\pi }-\gamma \right) ^{2}}\left(
\ln \frac{m_{\pi }+2\gamma }{m_{\pi }}-1\right) \right] \ \quad .
\label{sol3s13}
\end{equation}
Numerically we find
\begin{equation}
C_{0}^{(^{3}S_{1})}(m_{\pi })=-5.64\text{ fm}^{2}\quad ,\quad
D_{2}^{(^{3}S_{1})}(m_{\pi })=1.46\text{ fm}^{4}\quad ,\quad
C_{2}^{(^{3}S_{1})}(m_{\pi })=10.05\text{ fm}^{4}\quad .
\label{nsol3s1}
\end{equation}
The fit to the n-p scattering phase shifts is shown in Fig.~\ref{Fig:phaseshifts}.

%

\label{} 


\begin{references}

\bibitem{YHH}  S. Ying, W.C. Haxton, and E. M. Henley,
\Journal{\PRC}{45}{1982}{1992};
\Journal{\PRD}{40}{3211}{1989}.

\bibitem{KN}  K.\ Kubodera and S.\ Nozawa,
{\em Int.\ J.\ Mod.\ Phys.} {\bf E3},
101 (1994); Y.\ Kohyama and K.\ Kubodera, USC(NT)-Report-92-1, 1992,
unpublished.

\bibitem{ebcc}  S.D.\ Ellis and J.N.\ Bahcall,
\Journal{\NPA}{114}{636}{1968}.


\bibitem{adnc}  A.\ Ali and C.A.\ Dominguez, 
\Journal{\PRD}{12}{3673}{1975}.

\bibitem{HCLee}  H. C. Lee, 
\Journal{\NPA}{294}{473}{1978}.

\bibitem{GSW} S.L.\ Glashow, \Journal{\NP}{22}{579}{1961};
S.\ Weinberg, \Journal{\PRL}{19}{1264}{1967}; A.\ Salam,
in {\it Nobel Symposium}, {\bf No.\ 8}, ed.\ N.\ Svartholm
(Almquist and Wiksells, Stockholm, 1968).

\bibitem{Ardsma}  G.A. Aardsma {\it et al.}, 
\Journal{\PLB}{194}{321}{1987}.


\bibitem{SNO}  G.T. Ewan {\it et al.}, SNO proposal SNO-87-12, 1987.

\bibitem{BKN}  J.N.\ Bahcall, Kubodera, and S.\ Nozawa, 
\Journal{\PRD}{38}{1030}{1987}.

\bibitem{TKK}  N.\ Tatara, Y.\ Kohyama and K.\ Kubodera, 
\Journal{\PRC}{42}{1694}{1990}.

\bibitem{dksno}  M.\ Doi and K.\ Kubodera, 
\Journal{\PRC}{45}{1988}{1992}.


\bibitem{Weinberg1}S. Weinberg,
\Journal{\PLB}{251}{288}{1990};
\Journal{\NPB}{363}{3}{1991};
\Journal{\PLB}{295}{114}{1992}.

\bibitem{Bira}C. Ordonez and U. van Kolck, {\it Phys. Lett.} B {\bf 291},
459 (1992); C. Ordonez, L. Ray and U. van Kolck, {\it Phys. Rev. Lett.}
 {\bf 72}, 1982 (1994); {\it Phys. Rev. } C {\bf 53}, 2086 (1996);
U. van Kolck, {\it Phys. Rev. } C {\bf 49}, 2932 (1994).

\bibitem{Friara}J. Friar,
  {\tt nucl-th/9601012}; {\tt nucl-th/9601013};
  \Journal{\FBS}{99}{1}{1996};
  {\tt nucl-th/9804010};
  J.L. Friar, D. Huber, and U. van Kolck,
  \Journal{\PRC}{59}{53}{1999}.

\bibitem{Parka} T.-S.  Park, D.-P.  Min and M. Rho,
\Journal{\PRL}{74}{4153}{1995} ;
\Journal{\NPA}{596}{515}{1996}.


\bibitem{cohena}T.D. Cohen, 
\Journal{\PRC}{55}{67}{1997}.
D.R. Phillips and T.D. Cohen, 
\Journal{\PLB}{390}{7}{1997}.  
K.A. Scaldeferri, D.R. Phillips, C.W. Kao and T.D. Cohen,
\Journal{\PRC}{56}{679}{1997}.
S.R. Beane, T.D. Cohen and D.R. Phillips,
nucl-th/9709062;
D.R. Phillips, S.R. Beane and  T.D. Cohen,
{\it Annals Phys.} {\bf 263}, 255 (1998).
 
\bibitem{Sa96} M.J. Savage, 
\Journal{\PRC}{55}{2185}{1997}.


\bibitem{GPLa} G.P. Lepage, {\tt nucl-th/9706029},
Lectures at 9th Jorge Andre Swieca Summer School: 
Particles and Fields, Sao Paulo,
Brazil, Feb 1997.

\bibitem{LM} M. Luke and A.V. Manohar, {\it Phys. Rev.} D {\bf 55},
4129 (1997).

\bibitem{DR}D.B. Kaplan, M.J. Savage, and M.B. Wise, {\it Nucl. Phys.} 
 B {\bf 478}, 629 (1996).

\bibitem{KSW} D.B. Kaplan, M.J. Savage and M.B. Wise, 
\Journal{\PLB}{424}{390}{1998};
\Journal{\NPB}{534}{329}{1998}.


\bibitem{Kolck}U. van Kolck,
  \Journal{\NPA}{645}{273}{1999}

\bibitem{KSW2} D.B. Kaplan, M.J. Savage and M.B. Wise,
\Journal{\PRC}{59}{617}{1999}.

\bibitem{CGSSpol}J.W. Chen, H. W. Griesshammer, M. J. Savage and 
R. P. Springer,
\Journal{\NPA}{644}{221}{1999};\Journal{\NPA}{644}{245}{1999}.


\bibitem{Ccompt} J.W. Chen,
  {\tt nucl-th/9810021}, {\it to appear in Nucl. Phys.} A..

\bibitem{SSpv}M. J. Savage and R.P. Springer,
\Journal{\NPA}{644}{235}{1999}.

\bibitem{KSSWpv} D. B. Kaplan, M. J. Savage, R. P. Springer and 
M. B. Wise,
\Journal{\PLB}{449}{1}{1999}.

\bibitem{SSWst}
M. J. Savage, K. A. Scaldeferri, Mark B.Wise,
{\tt nucl-th/9811029}.

\bibitem{MehStew} T. Mehen and I.W. Stewart,
  {\tt nucl-th/9901064};
  {\tt nucl-th/9809095};
  \Journal{\PLB}{445}{378}{1999}.


\bibitem{GruSho} G. Rupak and N. Shoresh,
 {\tt nucl-th/9902077}.

\bibitem{Geg}J. Gegelia,
\Journal{\PLB}{429}{227}{1998};
{\tt nucl-th/9806028};
{\tt nucl-th/9805008};
{\tt nucl-th/9802038};

\bibitem{Int}A. K. Rajantie, {\it Nucl. Phys.} B {\bf 480}, 729 (1996). 

\bibitem{SteFurn} J.V. Steele and R.J. Furnstahl,
  \Journal{\NPA}{637}{46}{1998};
\Journal{\NPA}{645}{439}{1999}.

\bibitem{CohHan} T.D. Cohen and J.M. Hansen,
  \Journal{\PRC}{59}{13}{1999};
  {\tt nucl-th/9901065}.

\bibitem{Parkeft} T.-S. Park, K. Kubodera, D.-P. Min, and M. Rho,
\Journal{\PRC}{58}{637}{1998};
{\tt nucl-th/9807054}; {\tt astro-ph/9804144}.

\bibitem{Kong} X. Kong and F. Ravndal,
  {\tt nucl-th/9803046}; 
  \Journal{\PLB}{450}{320}{1999};
  {\tt nucl-th/9903523};
  {\tt nucl-th/9904066}.

\bibitem{Epel} E. Epelbaoum and U.G. Meissner, {\tt nucl-th/9902042};
E. Epelbaoum, W. Glockle, A. Kruger and  Ulf-G. Meissner,
\Journal{\NPA}{645}{413}{1999};
 E. Epelbaoum, W. Glockle and  Ulf-G. Meissner,
 \Journal{\PLB}{439}{1}{1998};
 E. Epelbaoum, W. Glockle and  Ulf-G. Meissner,
\Journal{\NPA}{637}{107}{1998}.
 
\bibitem{MSWsu} T. Mehen, I. W. Stewart and M.B. Wise,
{\tt hep-ph/9902370}.

\bibitem{PBB} D. R. Phillips, S. R. Beane and M. C. Birse,
  {\tt hep-ph/9810049}.

\bibitem{threebod} P.F. Bedaque, H.W. Hammer and U. van Kolck,
  \Journal{\PRL}{82}{463}{1999}; \Journal{\PRC}{58}{R641}{1998};
P.F. Bedaque and U. van Kolck,
 \Journal{\PLB}{428}{221}{1998}.

\bibitem{PMRsupp}  T.-S. Park, K. Kubodera, D.-P. Min and M. Rho,
 Talk presented by M. Rho at the 
  {\it Nuclear Physics with Effective Field Theory: 1999}
workshop, INT, University of Washington, Seattle, February 1999.
{\tt nucl-th/9904053}.

\bibitem{CRS} J.W. Chen, G. Rupak and M.J. Savage
  {\tt nucl-th/9902056}, {\it to appear in Nucl. Phys.} A.; 
  {\tt nucl-th/9905009}.

\bibitem{BMPV} S.R. Beane, M. Malheiro, D.R. Phillips and U. van Kolck
  {\tt nucl-th/9905023}.

\bibitem{KSte} D.B. Kaplan and J.V. Steele,
  {\tt nucl-th/9905027}.


\bibitem{deltas1} K.\ Abe {\it et al.}, \Journal{\PRL}{74}{346}{1995}.

\bibitem{deltas2} D.\ Adams {\it et al.}, \Journal{\PLB}{329}{399}{1994};
\Journal{\PLB}{339}{332}{1994}; \Journal {\PLB}{357}{248}{1995}.

\bibitem{deltas}  M.J.\ Savage and J.\ Walden,
\Journal{\PRD}{55}{5376}{1997}.

\bibitem{FHPY}  T. Frederico, E.M. Henley, S.J. Pollock, and S. Ying,
\Journal{\PRC}{46}{347}{1992}.

\bibitem{SAMPLE}  B. Mueller {\it et al.}, 
\Journal{\PRL}{78}{3824}{1997}.

\bibitem{strangetheory} M.J.\ Musolf {\it et al.},
\Journal{\PRep}{239}{1}{1994}.

\bibitem{SAMPLEII} Bates experiment \# 94-11 (M.\ Pitt and E.J.\ Beise, contacts). 

\bibitem{BCnopi} M.N.\ Butler and J.W.\ Chen, {\it in preparation}.

\bibitem{BCK} M.N.\ Butler, J.W.\ Chen, and X.\ Kong, {\it in preparation}.

\bibitem{VogelBeacom} P.\ Vogel and J.F.\ Beacom, {\tt hep-ph/9903554}.

\bibitem{Coon} S.A.\ Coon, {\tt nucl-th/9903033}.

\end{references}
\end{document}